\documentclass[rsi,superscriptaddress,twocolumn,amsmath,amssymb,reprent]{revtex4-1}
\usepackage{bm}% bold math
\usepackage{graphicx}%
\usepackage{dcolumn}% Align table columns on decimal point
\usepackage[colorlinks]{hyperref}
\hypersetup{allcolors=[RGB]{0,0,255}}
\usepackage[version=3]{mhchem} % Formula 
\usepackage{siunitx}
\begin{document}

%\linenumbers

\preprint{Manuscript ID: RSI19-AR-01381R}

\title{Development of High-Speed Ion Conductance Microscopy}

% \author{Shinji Watanabe}
% \email[]{wshinji@se.kanazawa-u.ac.jp}
% %\homepage[]{Your web page}
% \thanks{}
% %\altaffiliation{}
% \affiliation{Bio-AFM Frontier Research Center, Institute of Science and Engineering, Kanazawa University, Kakuma-machi, Kanazawa 920-1192, Japan}

% \author{Toshio Ando}
% \email[]{tando@se.kanazawa-u.ac.jp}
% %\homepage[]{Your web page}
% \thanks{}
% %\altaffiliation{}
% \affiliation{Bio-AFM Frontier Research Center, Institute of Science and Engineering, Kanazawa University, Kakuma-machi, Kanazawa 920-1192, Japan}

\author{Shinji Watanabe}
\email[]{wshinji@se.kanazawa-u.ac.jp}
\thanks{}
\affiliation{WPI Nano Life Science Institute (WPI-NanoLSI), Kanazawa University, Kakuma-machi, Kanazawa 920-1192, Japan}

\author{Satoko Kitazawa}
%\email[]{tando@se.kanazawa-u.ac.jp}
%\thanks{}
\affiliation{Department of Physics, Institute of Science and Engineering, Kanazawa University, Kakuma-machi, Kanazawa 920-1192, Japan}

\author{Linhao Sun}
%\email[]{tando@se.kanazawa-u.ac.jp}
%\thanks{}
\affiliation{WPI Nano Life Science Institute (WPI-NanoLSI), Kanazawa University, Kakuma-machi, Kanazawa 920-1192, Japan}

\author{Noriyuki Kodera}
%\email[]{tando@se.kanazawa-u.ac.jp}
%\thanks{}
\affiliation{WPI Nano Life Science Institute (WPI-NanoLSI), Kanazawa University, Kakuma-machi, Kanazawa 920-1192, Japan}

\author{Toshio Ando}
\email[]{tando@staff.kanazawa-u.ac.jp}
\thanks{}
\affiliation{WPI Nano Life Science Institute (WPI-NanoLSI), Kanazawa University, Kakuma-machi, Kanazawa 920-1192, Japan}

%! \date{\today} 必要であれば
\date{\today}% It is always \today, today,

\begin{abstract}
    Scanning ion conductance microscopy (SICM) can image the surface topography of specimens in ionic solutions without mechanical probe--sample contact.
    This unique capability is advantageous for imaging fragile biological samples but its highest possible imaging rate is far lower than the level desired in biological studies.
    Here, we present the development of high-speed SICM. The fast imaging capability is attained by a fast Z-scanner with active vibration control and pipette probes with enhanced ion conductance.
    By the former, the delay of probe Z-positioning is minimized to sub-\SI{10}{\micro s}, while its maximum stroke is secured at \SI{6}{\micro m}.
    The enhanced ion conductance lowers a noise floor in ion current detection, increasing the detection bandwidth up to \SI{100}{kHz}.
    Thus, temporal resolution 100-fold higher than that of conventional systems is achieved, together with spatial resolution around \SI{20}{nm}.
\end{abstract}

%! \pacs　必要であれば
\maketitle
%%%%%%%%%%%%%%%%%%%%%%%%%%%%%%%%%%%%%%%%%%%%%%%%%%%%%%%%%%%%%%%%%%%%%

%%%%%%%%%%%%%%%%%%%%%%%%%%%%%%%%%%%%%%%%%%%%%%%%%%%%%%%%%%%%%%%%%%%%%
%% Place any additional packages needed here.  Only include packages
%% which are essential, to avoid problems later. Do NOT use any
%% packages which require e-TeX (for example etoolbox): the e-TeX
%% extensions are not currently available on the ACS conversion
%% servers.
%%%%%%%%%%%%%%%%%%%%%%%%%%%%%%%%%%%%%%%%%%%%%%%%%%%%%%%%%%%%%%%%%%%%%

% \usepackage[version=3]{mhchem} % Formula 

% %mhchem:数式のフォーム
% \usepackage{amsmath}
% \usepackage{siunitx}
% %\usepackage{booktabs}
%%%%%%%%%%%%%%%%%%%%%%%%%%%%%%%%%%%%%%%%%%%%%%%%%%%%%%%%%%%%%%%%%%%%%
%% If issues arise when submitting your manuscript, you may want to
%% un-comment the next line.  This provides information on the
%% version of every file you have used.
%%%%%%%%%%%%%%%%%%%%%%%%%%%%%%%%%%%%%%%%%%%%%%%%%%%%%%%%%%%%%%%%%%%%%
%%\listfiles

%%%%%%%%%%%%%%%%%%%%%%%%%%%%%%%%%%%%%%%%%%%%%%%%%%%%%%%%%%%%%%%%%%%%%
%% Place any additional macros here.  Please use \newcommand* where
%% possible, and avoid layout-changing macros (which are not used
%% when typesetting).
%%%%%%%%%%%%%%%%%%%%%%%%%%%%%%%%%%%%%%%%%%%%%%%%%%%%%%%%%%%%%%%%%%%%%
\newcommand*\mycommand[1]{\texttt{\emph{#1}}}

\section{Introduction}

\begin{figure}[!t]
\includegraphics{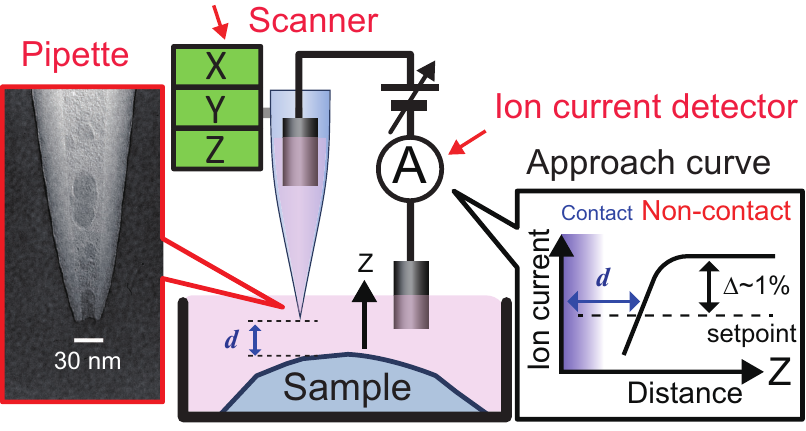}
\caption{\label{FIG1}
%Working principle of SICM. The electrolyte-filled tip with a nanopore at its distal end (see the scanning electron microscope image) is mounted on the scanner. The ion current through the nanopore induced by the application of bias voltage between the two Ag/AgCl electrodes is measured by the ion current detector. The measured ion current, which is dependent on the tip--surface separation, is used as a Z-position control signal. $\Delta _{\textrm{length}}$ is described in the main text.
Working principle of SICM.
%The electrolyte-filled pipette tip with a nanopore at its end (see the scanning electron micrograph in the left panel) is mounted on the scanner.
%![Revision]
The electrolyte-filled pipette with a nanopore at its end (see the transmission electron micrograph in the left panel) is mounted on the scanner.
% \textcolor{red}{
% Although the inside of the pipette appears to be filled with something, this happens during procedures for TEM imaging.}
The ion current through the nanopore generated by the application of bias voltage between the two Ag/AgCl electrodes is measured by the ion current detector. The measured ion current, which is dependent on the tip--surface separation $d$, is used as a pipette Z-position control signal.
}
\end{figure}

Tapping mode atomic force microscopy (AFM)~\cite{hansma1994tapping} has been widely used to visualize biological samples in aqueous solution with high spatial resolution.
However, when the sample is very soft, like eukaryotic cell surfaces, the intermittent tip-sample contact significantly deforms the sample and hence blurs its image~\cite{zhang2012scanning,ushiki2012scanning,ando2018high}.
Moreover, when the sample is extremely fragile, it is often seriously damaged~\cite{seifert2015comparison,ando2018high}.
SICM was invented to overcome this problem~\cite{hansma1989scanning}.
SICM uses as a probe an electrolyte-filled pipette having a nanopore at the tip end, and measures an ion current that flows between an electrode inside the pipette and another electrode in the external bath solution.
The ionic current resistance between the pipette tip and sample surface (referred to as the access resistance) increases when the tip approaches the sample. This sensitivity of access resistance to the tip--sample distance enables imaging of the sample surface without mechanical tip-sample contact~\cite{del2014contact,thatenhorst2014effect} (Fig. \ref{FIG1}).
To improve fundamental performances of SICM, several devices have recently been introduced, including a technique to control the pore size of pipettes~\cite{steinbock2013controllable,xu2017controllable,sze2015fine} and a feedback control technique based on tip--sample distance modulation~\cite{pastre2001characterization,li2014phase,li2015amplitude}.
Moreover, the SICM nanopipette has recently been used to measure surface charge density~\cite{mckelvey2014surface,mckelvey2014bias,page2016fast,perry2015simultaneous,perry2016surface,klausen2016mapping,fuhs2018direct} and electrochemical activity~\cite{kang2017simultaneous,takahashi2012topographical} as well as to deliver species~\cite{bruckbauer2002writing,babakinejad2013local,page2017quantitative,takahashi2011multifunctional}.
Thus, SICM is now becoming a useful tool in biological studies, especially for characterizing single cells with very soft and fragile surfaces~\cite{page2017multifunctional}.

However, the imaging speed of SICM is low; it takes from a few minutes to a few tens of minutes to capture an SICM image, which is in striking contrast to AFM.
High-speed AFM is already established~\cite{ando2008high} and has been used to observe a variety of proteins molecules and organelles in dynamic action~\cite{ando2014filming}.
The slow performance of SICM is due mainly to a low signal-to-noise ratio (SNR) of ion current sensing, resulting in its low detection bandwidth (and hence low feedback bandwidth). Moreover, the low resonant frequency of the Z-scanner also limits the feedback bandwidth.

\hypertarget{}{When} the vertical scan of the pipette towards the sample is performed with velocity $v_{\textrm{z}}$, the time delay of feedback control $t_{\textrm{delay}}$ causes an overshoot for the vertical scan by $t_{\textrm{delay}} \times v_{\textrm{z}}$. This overshoot distance should be smaller than the closest tip--sample distance ($d_{\textrm{c}}$) to be maintained during imaging (see the approach curve in Fig. \ref{FIG1}).
That is,

\begin{equation}
\label{fall_v}
v_{\textrm{z}} \leq \frac{d_{\textrm{c}}}{t_{\textrm{delay}}}.
\end{equation}

%where $\Delta t_{\textrm{delay}}$ represents the delay of the tip-position control. $\Delta I_{\textrm{length}}$ is related to the pipette geometry~\cite{rheinlaender2009image,del2014contact,korchev1997specialized}: the nanopore radius $r_{\textrm{a}}$, the cone angle, and the outer diameter of the tip, and an empirical value $\Delta I_{\textrm{length}}$ = $2r_{\textrm{a}}$ allows us to roughly estimate the attainable fall velocity. In SICM setup, $\Delta t_{\textrm{delay}}$ can be roughly estimated from the addition of the inverse of the mechanical resonance frequency of the Z-nanopositioner $f_{\textrm{res}}$ and the inverse of the transimpedance amplifier bandwidth $B_{\textrm{id}}$; thus, $\Delta t_{\textrm{delay}} \sim f_{\textrm{res}}^{-1} + B^{-1}$. A typical imaging condition of a commercially available SICM apparatus is $r_{\textrm{a}} \sim 25$ nm, $f_{\textrm{res}} \sim 1$ kHz, and $B \sim 1$ kHz, yielding a fall velocity of 25 nm/ms. With this value of the fall rate, SICM takes $\sim$1000 seconds to capture topographic images (100 $\times$ 100 pixels, and 100 ms to generate one-pixel data in average) of a sample with a surface roughness of 1 $\mu$m~\cite{zhang2005scanning,tanaka2015time,novak2009nanoscale,gesper2015long,nashimoto2015nanoscale}. Thus, morphological changes that occur with the span of a few minutes cannot be visualized using SICM with an acceptable image quality.

An appropriate size of $d_{\textrm{c}}$ is related to the pipette geometry, such as the tip aperture radius $r_{\textrm{a}}$, the cone angle $\theta_{\textrm{c}}$, and the outer radius of the tip $r_{\textrm{o}}$~\cite{rheinlaender2009image,del2014contact,korchev1997specialized}, but $d_{\textrm{c}} \approx 2 r_{\textrm{a}}$ is typically used to achieve highest possible resolution. The size of $t_{\textrm{delay}}$ can be roughly estimated from the resonant frequency of the Z-scanner $f_{\textrm{z}}$ and the bandwidth of ion current detection $B_{\textrm{id}}$, as $t_{\textrm{delay}} \approx 1/f_{\textrm{z}} + 1/B_{\textrm{id}}$.
In typical SICM setups, the values of these parameters are $r_{\textrm{a}} \approx$ \SI{20}{nm}, $f_{\textrm{z}} \approx$\SI{1}{kHz} and $B_{\textrm{id}} \approx$ \SI{1}{kHz}, yielding $v_{\textrm{z}} <$ \SI{20}{\micro m/s}. In the representative SICM imaging mode referred to as the hopping mode~\cite{novak2009nanoscale}, the tip-approach and retract cycle is repeated for a distance (hopping amplitude) of slightly larger than the sample height, $h_{\textrm{s}}$.
%For example, when $v_{\textrm{z}} =$ \SI{20}{\micro m/s} is used for the sample with $h_{\textrm{s}} \sim$ \SI{1}{\micro m}, it takes at least $>$\SI{50}{ms}, which depends on the retraction speed, to acquire pixel height information, corresponding to a imaging acquisition time of $\sim$\SI{10}{min} for 100 $\times$ 100 pixel resolution~\cite{novak2014imaging}. When the pipette retraction speed can be set at larger than the $v_{\textrm{z}}$, the imaging acquisition time can be improved but not much.
For example, when $v_{\textrm{z}} =$ \SI{20}{\micro m/s} is used for the sample with $h_{\textrm{s}} \approx$ \SI{1}{\micro m}, it takes at least $>$ \SI{50}{ms} for pixel acquisition, which depends on the retraction speed. This pixel acquisition time corresponds to an imaging acquisition time longer than 8.3 min for 100 $\times$ 100 pixel resolution~\cite{novak2014imaging}.
%!8.3は　50ms (acq/pixel)*100*100/60 = x min.
%!Q（AFMのように同じライン上を往復してイメージングしないのですね？）
%!A そのとおりです。SICMだとこれが一般的ですね。
When the pipette retraction speed can be set at much larger than the approach speed $v_{\textrm{z}}$, the imaging acquisition time can be improved but not much.

\hypertarget{FIG1}{}
Several groups have attempted to increase $v_{\textrm{z}}$~\cite{shevchuk2012alternative,novak2014imaging,jung2015closed,kim2015alternative,li2014phase,li2015phase}.
One of approaches used is to mount a shear piezoactuator with a high resonant frequency (but with a small stroke length) on the Z-scanner and this fast piezoactuator is used as a `brake booster'~\cite{shevchuk2012alternative,novak2014imaging}; that is, this piezoactuator is activated only in the initial retraction phase where the tip is in close proximity to the surface.
This method could cancel an overshooting displacement and therefore increase $v_{\textrm{z}}$ by 10-fold.
Another approach is to increase $B_{\textrm{id}}$ by the improvement of the SNR of current signal detection with the use of a current-source amplification scheme~\cite{kim2015alternative} or by the use of AC bias voltage between the electrodes (the AC current in phase with the AC bias voltage is used as an input for feedback control)~\cite{li2014phase,li2015phase}.
This bias voltage modulation method is further improved by capacitance compensation~\cite{li2015amplitude}.
%However, the improvement of speed performance of SICM by these methods are limited to a few times at most. Thus far, no efforts to extensively increase both $f_{\textrm{z}}$ and $B_{\textrm{id}}$ have been made.
The improvement of SICM speed performance by these methods is however limited to a few times at most.
Very recently, two studies demonstrated fast imaging of live cells with the use of their high-speed SICM (HS-SICM) systems~\cite{ida2017high,simeonov2019high}.
However, one of these studies used temporal tip--sample contact to alter hopping amplitude~\cite{ida2017high}, while the other used pipettes with $r_{\textrm{a}}$ = 80--\SI{100}{nm} and abandoned optical observation of the sample~\cite{simeonov2019high}.
Note that in SICM the temporal resolution has a trade-off relationship with the spatial resolution, as in the cases of other measurement techniques.
Thus far, no attempts have been made to increase both $f_{\textrm{z}}$ and $B_{\textrm{id}}$ extensively, without compromise of the spatial resolution and non-contact imaging capability of SICM. 

%Herein, we present the development of a high-speed ion conductance microscopy (HS-ICM) and first demonstrate high-speed images with high spatial resolution using HS-ICM. An image rate was revolutionary improved from $\sim$1000 (standard SICM) to \SI{10}{s/frame} or less. The two key developments, (i) a tip-scan-type high-speed nanopositioner~\cite{watanabe2017high} and its tip-position control techniques and (ii) an SNR-enhanced tip with an ion concentration gradient, allow us to overcome the current SICM temporal resolution. The improved $f_{\textrm{res}}$ of our nanopositioner and undesired vibration-control techniques provide a time response of $\sim$\SI{3}{\micro s}, corresponding to a 100-fold improvement in the delay ($\sim$1--\SI{3}{ms} in a standard SICM). The improved SNR is a factor of $\sim$8, pushing the bandwidth of the ion current detector from 1 to \SI{100}{kHz} or more. We found that the mechanism of this enhancement of SNR can be explained by our finite-element method (FEM) modeling based on coupled Poisson--Nernst--Planck equations and it is valid for high-speed scanning. We demonstrate that HS-ICM can capture biological structures with sub-\SI{10}{nm} scale within a few seconds per frame. This indicates that HS-ICM will be a promising tool to visualize the dynamics of biological systems in a liquid environment with high spatial resolution.

%! revision rates -> rate
Here, we report the development of HS-SICM and demonstrate its high-speed and high resolution imaging capability.
The image rate was improved by a factor of $\sim$100 or slightly more.
This remarkable enhancement in speed was achieved by two improved performances:(i) fast pipette positioning achieved with the developed fast scanner and vibration suppression techniques and (ii) an enhanced SNR of current detection by reduction of the ionic resistance arising from the inside of the pipette (referred to as the pipette resistance).
The improved $f_{\textrm{z}}$ of the Z-scanner resulted in a mechanical response time of $\sim$\SI{3}{\micro s}, corresponding to a $\sim$100-fold improvement over conventional SICM systems. The SNR of current detection was improved by a factor of $\sim$8, enhancing $B_{\textrm{id}}$ from 1 to \SI{100}{kHz} or slightly higher.
The HS-SICM system was demonstrated to be able to capture topographic images of low-height biological samples at 0.9--\SI{5}{s/frame} and live cells at 20--\SI{28}{s/frame}.
These high imaging rate performances are compatible with spatial resolution of 15--\SI{23}{nm}. 

%\iffigure
%\fi

\section{Results and Discussion}

%\subsection{Strategy for High-Speed Scan}
\subsection{Strategy towards HS-SICM}

%A hopping scan mode~\cite{novak2009nanoscale} is frequently used for imaging a bumpy surface in SICM. Thus, improving the scan speed of the hopping mode has a significant impact on a practical use of SICM and this is our aim in this paper. In the hopping mode, the tip approach and retract with a certain distance amplitude (hopping amplitude) is needed to produce one-pixel topographic data. Thus, the scan speed determining the temporal resolution is limited by the fall velocity of the tip, as described in Eq. (\ref{fall_v}). Because the attainable fall velocity increases with decreasing $\Delta t_{\textrm{delay}} \sim f_{\textrm{res}}^{-1} + B^{-1}$, improvements in both $f_{\textrm{res}}$ and $B_{\textrm{id}}$ are required. No systematic studies have been carried out to improve the both $f_{\textrm{res}}$ and $B_{\textrm{id}}$.

The speed of pipette approach towards the sample ($v_{\textrm{z}}$) is limited, as expressed by Eq. (\ref{fall_v}).
As $v_{\textrm{z}}$ depends on $f_{\textrm{z}}$ and $B_{\textrm{id}}$, the improvement on both $f_{\textrm{z}}$ and $B_{\textrm{id}}$ is required to achieve HS-SICM.
To increase $f_{\textrm{z}}$, we need a fast Z-scanner for displacing the pipette along its length. Note that all commercially available Z-scanners for SICM have $f_{\textrm{z}} <$ \SI{10}{kHz}.
Besides, we need to establish a method to mount the pipette ($\sim$\SI{15}{mm} in length) to the Z-scanner in order to minimize the generation of undesirable vibrations of  the pipette. 
We previously developed a fast XYZ scanner with $f_{\textrm{z}} \approx$ \SI{100}{kHz}, a resonant frequency of $\sim$\SI{2.3}{kHz} in the XY directions, and stroke distances of $\sim$\SI{6}{\micro m} and $\sim$\SI{34}{\micro m} for Z and XY, respectively~\cite{watanabe2017high}.
In this study, we further improved the dynamic response of this fast scanner.
%Considering this high $f_{\textrm{z}}$, we need to increase $B_{\textrm{id}}$ to the level of $\sim$\SI{100}{kHz}. As the ion current change caused by an altered tip-sample distance is generally small ($\sim$\SI{1}{pA}), the current signal noise largely limits $B_{\textrm{id}}$.
Considering this high $f_{\textrm{z}}$ with improved dynamic response, we need to increase $B_{\textrm{id}}$ to the level of $\sim$\SI{100}{kHz}.
As the ion current change caused by an altered tip-sample distance is generally small ($\sim$\SI{1}{pA}), the current signal noise largely limits $B_{\textrm{id}}$.
At a high frequency regime ($>$ \SI{10}{kHz}), the dominant noise source is the interaction between the amplifier's current noise and the total capacitance at the input~\cite{rosenstein2012integrated,levis1993use,rosenstein2013single}.
Therefore, we have to lower the total capacitance and increase the current signal to achieve $B_{\textrm{id}} \approx$ \SI{100}{kHz}, without increasing the pipette pore size.

\begin{figure*}[!t]
\includegraphics{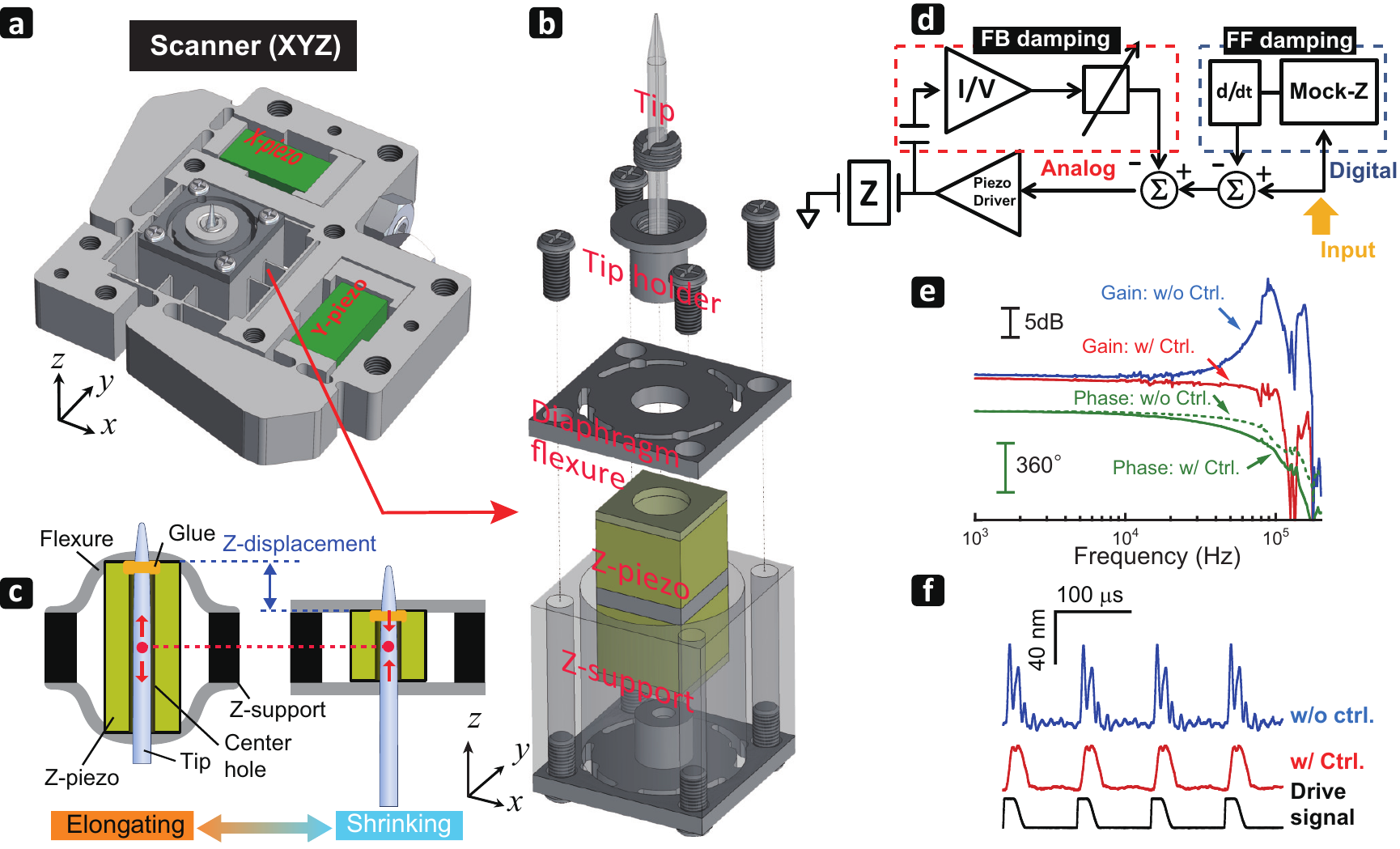}
\caption{\label{FIG2}
%(\textbf{a}) Assembly drawing of the XYZ-scanner used in this study. A lateral scan was performed with the X- and Y-piezoactuators via a leverarm-displacement amplification scheme. (\textbf{b}) Exploded view of the Z-scanner. (\textbf{c}) Assembly drawing showing the inertia balance design of Z-scanner. (\textbf{d}) Schematic of feedforward and feedback damping control methods. (\textbf{e}) Transfer functions of Z-scanner. The blue and red solid lines indicate the gain signals measured without and with the use of the damping control methods, respectively. The green broken and solid lines represent the phase for the cases without and with the controls. (\textbf{f}) Time domain responses of the Z-scanner to the application of a square-like waveform to the Z-piezoactuator. The blue and red solid lines represent the displacement of the Z-scanner for the cases without and with the use of the damping control methods, respectively.
XYZ-scanner used in this study. (\textbf{a}) Assembly drawing of the scanner.
For the lateral scan, the displacements of X- and Y-piezoactuators are magnified via a lever arm  amplification scheme.
(\textbf{b}) Exploded view of the Z-scanner.
(\textbf{c}) Assembly drawing showing the inertia balance design used for the Z-scanner.
(\textbf{d}) Schematic of FF and FB damping control methods. (\textbf{e}) Transfer functions (frequency response) of Z-scanner. The blue and red lines indicate the gain signals measured without and with the damping control methods, respectively.
The green broken and solid lines indicate the phase signals measured without and with the damping control methods, respectively.
%(\textbf{f}) Time domain responses of the Z-scanner to the application of a square-like waveform voltage to the Z-piezoactuator.
(\textbf{f}) Time domain responses of the Z-scanner in the application of a square-like waveform voltage to the input (black line).
%The blue and red lines show the measured Z-scanner displacements for the cases without and with the use of the damping control methods, respectively.
The blue and red lines show the measured Z-scanner displacements without and with the use of the damping control methods, respectively.
}
\end{figure*}

%\subsection{High-Speed Z-Nanopositioner}
\subsection{High-speed Z-scanner}

%! revision Fig. Figure -> Fig.
The structure of our fast scanner developed is shown in Fig. \ref{FIG2}a-c.
A key mechanism for minimizing unwanted Z-scanner vibrations is momentum cancellation; the hollow Z-piezoactuator is sandwiched with a pair of identical diaphragm-like flexures, so that the center of mass of the Z-piezoactuator hardly changes during its fast displacement (\href{https://aip.scitation.org/doi/suppl/10.1063/1.5118360/suppl_file/rsi_sup_%28sw%29191112%28clear%29.pdf}{Supplementary material}, SI 1).
The pipette is mechanically connected only with the top flexure through being glued to the top clamp.
%!revised 191116, noticeably undesirable -> noticeable
Thanks to these designs, no noticeable resonance peaks are induced except at the resonant frequency of the Z-piezoactuator (Fig. \ref{FIG2}e).
We achieved a product value of \SI{6}{\micro m} (maximum displacement) $\times$ \SI{100}{kHz} (resonant frequency) in this Z-scanner, which exceeds more than 10-fold the value of conventional designs of SICM Z-scanner, \SI{25}{\micro m} $\times$ 1--\SI{2}{kHz}.
In the present study, we further improved the dynamic response of Z-scanner.
The sharp resonant peak shown in Fig. \ref{FIG2}e (blue line) induces unwanted vibrations. In fact, the application of a square-like-waveform voltage to the Z-scanner (black line in Fig. \ref{FIG2}f) generated an undesirable ringing displacement of the Z-scanner (blue line in Fig. \ref{FIG2}f).
To damp this ringing, we developed feedforward (FF) and feedback (FB) control methods (Fig. \ref{FIG2}d).
The FF control system was implemented in field-programmable-gate-array (FPGA). The gain controlled output signal from a mock Z-scanner (an electric circuit) with a transfer function similar to that of the real Z-scanner was first differentiated and then subtracted from the signal input to the Z-piezodriver~\cite{kodera2005active}.
%Although this method was effective in reducing the Q-factor of the Z-scanner, a slight difference between the transfer functions of mock and real Z-scanners kept residual vibrations.
%To remove the residual vibrations, the FB control implemented in an analog circuit was added.
%　（データは示さないのか？）
Although this method was effective in reducing the Q-factor of the Z-scanner, the drift behavior of the transfer function of real Z-scanner would affect the reduced Q-factor during long-term scanning.
To suppress the drift effect, the FB control implemented in an analog circuit was added as follows.
The velocity of Z-scanner displacement was measured using the transimpedance amplifier via a small capacitor of $\sim$\SI{1}{pF} positioned near the Z-scanner.
The gain-controlled velocity signal was subtracted from the output of the FF controller.
In this way, too fast movement of the Z-scanner was prevented~\cite{kageshima2006wideband}, resulting in nearly complete damping of unwanted vibrations, as shown with the red lines of Figs. \ref{FIG2}e and \ref{FIG2}f.
%Thus, the open-loop response time of Z-scanner, $Q/\pi f_{\textrm{z}}$, was improved from $\sim$\SI{12}{\micro s} to \SI{3}{\micro s} (critical damping).
%Note that a delay of $\sim$\SI{3.5}{\micro s} seen in Fig. \ref{FIG2}f between the drive signal (black line) and the Z-scanner displacement (blue and red lines) is due to the latency originating from the FPGA circuit (\SI{0.5}{\micro s}), the piezodriver (\SI{1}{\micro s}), and the laser vibrometer used (\SI{2}{\micro s}) (Supplementary material, SI 1, Fig. S2).
%The FF/FB damping control was also applied to the XY scanners to improve their dynamic response (Supplementary material, SI 1, Fig. S3).%
Thus, the open-loop response time of Z-scanner, $Q/\pi f_{\textrm{z}}$, was improved from $\sim$\SI{18.5}{\micro s} to \SI{1.8}{\micro s} (critical damping).
Note that measured Z-scanner displacements (blue nad red lines) include the latency of the laser vibrometer used (\SI{2}{\micro s}) (\href{https://aip.scitation.org/doi/suppl/10.1063/1.5118360/suppl_file/rsi_sup_%28sw%29191112%28clear%29.pdf}{Supplementary material}, SI 1, Fig. S2).
The FF/FB damping control was also applied to the XY scanners to improve their dynamic response (\href{https://aip.scitation.org/doi/suppl/10.1063/1.5118360/suppl_file/rsi_sup_%28sw%29191112%28clear%29.pdf}{Supplementary material}, SI 1, Fig. S3).
\subsection{Enhancement of SNR with Salt Concentration Gradient}

\begin{figure}[!tb]
\includegraphics{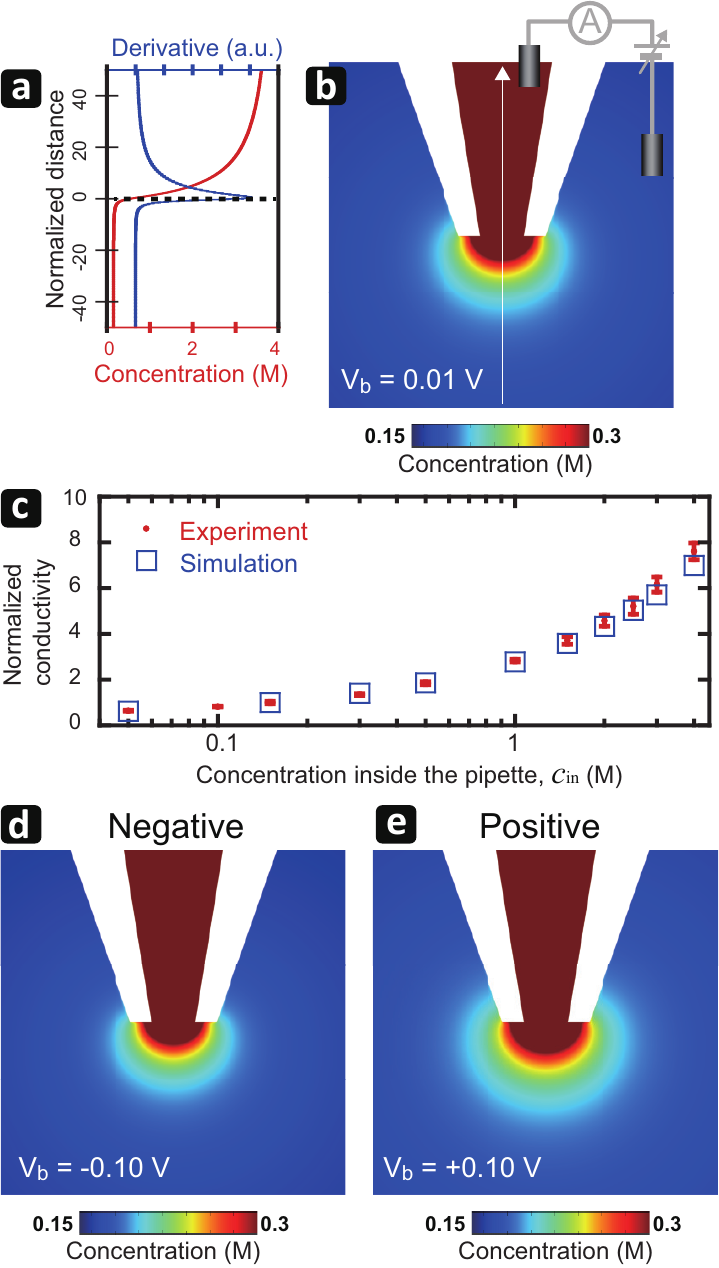}
\caption{\label{FIG4}
%Line (\textbf{a}) and overall (\textbf{b}) spatial profiles of average ion concentration of K$^+$ and Cl$^-$ at $c_{\textrm{p}}$ = \SI{4}{M}, $c_{\textrm{in}}$ = \SI{0.15}{M} and $V_{\textrm{b}}$ = \SI{0.01}{V} were obtained using FEM simulations for a tip with a surface charge density of \SI{-10}{mC/m^2}. (\textbf{a}) The line profile along the white arrow in (\textbf{b}). Vertical axis of (\textbf{a}) represents the distance normalized by the tip aperture diameter. (\textbf{c}) Enhancement of tip conductivity by the concentration gradient. Vertical axis of (\textbf{c}) represents the enhancement factor of the tip conductivity with respect to that obtained at an inner salt concentration of \SI{0.15}{M}.
Spatial distribution of average concentration of K$^+$ and Cl$^-$ obtained by FEM simulation. %for the case of $c_{\textrm{p}}$ = \SI{4}{M}, $c_{\textrm{in}}$ = \SI{0.15}{M}, and $V_{\textrm{b}}$ = \SI{0.01}{V}.
The surface charge density of the tip was set at \SI{-10}{mC/m^2}. (\textbf{a}) Average ion concentration profile (red) and its derivative (blue) along the white arrow shown in (\textbf{b}). 
$c_{\textrm{in}}$ (KCl) = \SI{4}{M}, $c_{\textrm{out}}$ (KCl) = \SI{0.15}{M}  and $V_{\textrm{b}}$ = \SI{0.01}{V} were used.
%The vertical axis represents the Z-distance from the tip aperture normalized with the tip aperture diameter. 
%! [Revision]追加
The vertical axis represents the Z-distance from the tip aperture normalized with the tip aperture diameter; the tip aperture position is zero as indicated by the broken line. 
(\textbf{b}) Spatial distribution of average ion concentration under the same conditions as (\textbf{a}). (\textbf{c}) Enhancement of tip conductivity by ICG.
The vertical axis represents the enhancement factor of the tip conductivity with respect to the ion conductivity at $c_{\textrm{in}}$ = \SI{0.15}{M}. (\textbf{d, e}) Spatial distributions of average ion concentration at $V_{\textrm{b}}$ = \SI{-0.1}{V}(\textbf{d}) and $V_{\textrm{b}}$ = \SI{0.1}{V}(\textbf{e}) for $c_{\textrm{in}}$ (KCl) = \SI{4}{M} and $c_{\textrm{out}}$ (KCl) = \SI{0.15}{M}.
}
\end{figure}

We describe here a method to improve $B_{\textrm{id}}$ by increasing the SNR of current signal sensing. In the frequency region $>$ \SI{10}{kHz}, the dominant noise source of the ion current detector is the total capacitance at the transimpedance amplifier input, $\Sigma \textrm{C}$~\cite{rosenstein2012integrated,levis1993use}:

%Now, $\Delta t_{\textrm{delay}}$ is governed by $B_{\textrm{id}}$. We describe here the method to improve $B_{\textrm{id}}$ by increasing the SNR. The dominant noise source of the ion current detector in the frequency region $>$ \SI{10}{kHz} is a total capacitance at the transimpedance amplifier input $\Sigma \, \textrm{C}$ ~\cite{rosenstein2012integrated,levis1993use}:
\begin{equation}
I_{\textrm{RMS}} \propto B_{\textrm{id}}^{3/2} \Sigma \, \textrm{C},
\end{equation}
% where $I_{\textrm{RMS}}$ represents a root-mean-square current noise.
% The electrode-wiring ($\sim$\SI{3}{pF}) and the pipette capacitance $C_\textrm{p} \sim$\SI{1}{pF} dominate the total capacitance.
where $I_{\textrm{RMS}}$ represents a root-mean-square current noise.
The electrode-wiring and the pipette capacitance $C_\textrm{p}$ dominate the total capacitance.
As the $C_{\textrm{p}}$ derives from the part of the pipette immersed in solution, thicker wall pipettes are useful in reducing $C_{\textrm{p}}$.
We used quartz capillaries with a wall thickness of 0.5--\SI{0.7}{mm}.
%The value of $\Sigma \textrm{C}$ in our setup was estimated to be $\sim$\SI{5}{pF} (Supplementary material, Fig. S9)%! 引用する順序にすべき
The total capacitance in our setup was estimated to be $\sim$\SI{5}{pF} (\href{https://aip.scitation.org/doi/suppl/10.1063/1.5118360/suppl_file/rsi_sup_%28sw%29191112%28clear%29.pdf}{Supplementary material}, SI 2), yielding $I_{\textrm{RMS}}$ $\sim$ \SI{8}{pA} at $B_{\textrm{id}}$ = \SI{100}{kHz} (although $I_{\textrm{RMS}}$ was $\sim$\SI{1.2}{pA} at $B_{\textrm{id}}$ = \SI{10}{kHz}), which was still too large.
Then, we decided to increase the ion current to improve the SNR further.
Since the bias voltage ($V_{\textrm{b}}$) larger than a typical value of $\pm$ \SI{0.5}{V} induces an unstable ion current~\cite{clarke2012pipette}, we need to reduce the pipette resistance ($R_{\textrm{p}}$).
The ion current $I_{\textrm{i}}$ through the pipette opening is approximately described as
\begin{equation}
I_{\textrm{i}}(d) = \frac{V_{\textrm{b}}}{R_{\textrm{a}}(d) + R_{\textrm{p}}},
\label{I(d)}
\end{equation}
where $d$ is the tip-surface distance and $R_{\textrm{a}}$ is the access resistance that depends on $d$~\cite{edwards2009scanning}.
In Eq. \ref{I(d)}, the surface charge-dependent ion current rectification in the pipette is not considered~\cite{wei1997current}.
$R_{\textrm{p}}$ is usually $\sim$100-times larger than $R_{\textrm{a}}$ even at $d \approx d_{\textrm{c}}$, and therefore, the reduction of $R_{\textrm{p}}$ directly increases $I_{\textrm{i}}$.
To reduce $R_{\textrm{p}}$, we examined the ion concentration gradient (ICG) method; a pipette back-filled with a high salt solution is immersed in a low salt solution.
%ここから190129
Since the pipette opening is very small, a concentration gradient is expected to be formed only in the close vicinity of the pipette opening.
Although several studies have been performed on ICG from the viewpoint of its effect on the ion current rectification in nanopores~\cite{cao2011concentration,deng2014effect,yeh2014tuning}, it is unclear whether or not the ICG method is really useful and applicable to SICM, as the physiological salt concentration used in the external bath solutions is relatively high.
%! [Revise] fine -> finite
% To check this issue, we first performed a fine element method (FEM) simulation using the coupled Poisson--Nernst--Planck (PNP) equations that have been widely adopted to study the transport behavior of charged species~\cite{bazant2009towards,klausen2016mapping,perry2016characterization}.
To check this issue, we first performed a finite element method (FEM) simulation using the coupled Poisson--Nernst--Planck (PNP) equations that have been widely adopted to study the transport behavior of charged species~\cite{bazant2009towards,klausen2016mapping,perry2016characterization}.
Full details of our PNP simulation setup are described in \href{https://aip.scitation.org/doi/suppl/10.1063/1.5118360/suppl_file/rsi_sup_%28sw%29191112%28clear%29.pdf}{Supplementary material}, SI 3 and Methods.
Figures \ref{FIG4}a,b show a FEM simulation result obtained for the spatial profile of total ion concentration $(c_{\textrm{K}^+} + c_{\textrm{Cl}^-})/2$, when \SI{4}{M} KCl and physiological \SI{0.15}{M} KCl solutions were used for the inside and outside of the pipette, respectively.
As seen there, the region of ICG is confined in a small volume around the pipette opening, while the outside salt concentration is maintained at $\sim$\SI{0.23}{M} and $<$ \SI{0.17}{M} in the regions distant from the opening by $\sim$2$r_{\textrm{a}}$ and $>$ $4r_{\textrm{a}}$, respectively.
Figure \ref{FIG4}c (square plots) shows a simulation result for changes of ion conductance 1/$R_{\textrm{p}}$ when the KCl concentration inside the pipette ($c_{\textrm{in}}$) was altered, while the outside bulk KCl concentration ($c_{\textrm{out}}$) was kept at \SI{0.15}{M}.
This result was very consistent with that obtained experimentally (red plots in Fig. \ref{FIG4}c).
The value of $1/R_{\textrm{p}}$ at $c_{\textrm{in}}$ = \SI{4}{M} was  $\sim$8-times larger than that at $c_{\textrm{in}}$ = \SI{0.15}{M}.
%! revision c_in <-> c_out
We also confirmed that the conditions of $c_{\textrm{in}}$ = \SI{4}{M} and $c_{\textrm{out}}$ = \SI{0.15}{M} generate a steady current with $|V_{\textrm{b}}| <$ \SI{0.5}{V}, and hence, allow stable SICM measurements for \SI{7.5}{nm} $\leq r_{\textrm{a}} \leq$ \SI{25}{nm}.
Note that the high KCl concentration region can be confined to a smaller space when a negative bias voltage is used because of an ion current rectification effect of the negatively charged pipette (Fig. \ref{FIG4}d, e).
% （どこにも示していないけど）
%! 示す必要は無いように思います。

\begin{figure}[!tb]
\includegraphics{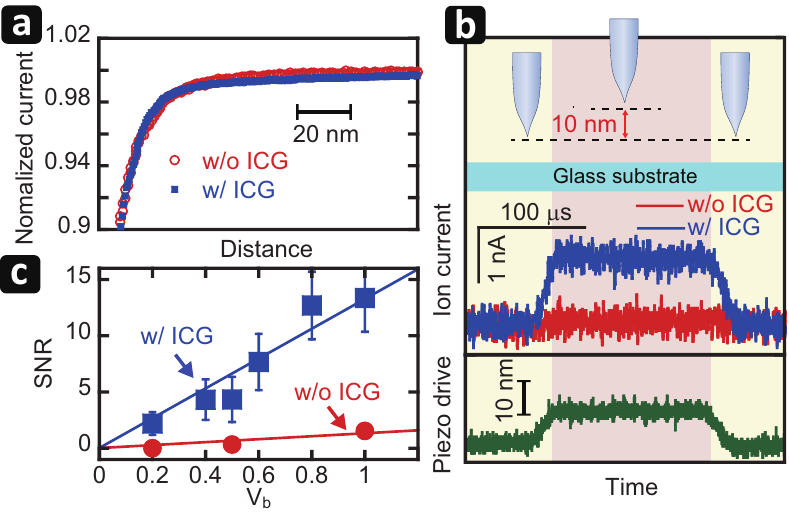}
\caption{\label{FIG5}
Enhancement of ion current response by ICG. (\textbf{a}) Approach curves with (blue) and without (red) ICG method. (\textbf{b}) Dynamic response of ion current at $V_{\textrm{b}}$ = \SI{0.5}{V} when the tip is vertically moved (shown in green) in close proximity to the glass surface, and its dependence on the use (shown in blue) and non-use (shown in red) of ICG method. (\textbf{c}) Increase of SNR of ion current measurement with increasing $V_{\textrm{b}}$ and its dependence on the use (blue) and non-use (red) of ICG method.
}
\end{figure}

To confirm the SNR enhancement of $I_{\textrm{i}}$ by ICG formed by the use of $c_{\textrm{in}}$ = \SI{4}{M} and $c_{\textrm{out}}$ = \SI{0.15}{M}, we measured the dynamic responses of $I_{\textrm{i}}$ to quick change of $d$ under $V_{\textrm{b}}$ = \SI{0.5}{V}, in the presence and absence of IGC.
To measure the responses, the pipette with $r_{\textrm{a}}$ = \SI{10}{nm} was initially positioned at a Z-point showing 5$\%$ reduction of $I_{\textrm{i}}$ (see Fig. \ref{FIG5}a).
Then, the pipette was quickly retracted by \SI{10}{nm} within \SI{14}{\micro s}, and after a while quickly approached by \SI{10}{nm} within \SI{14}{\micro s}(Fig. \ref{FIG5}b, Top), by the application of a driving signal with a rectangle-like waveform (Fig. \ref{FIG5}b, Bottom) to the developed Z-scanner.
The ion current responses measured using the transimpedance amplifier with $B_{\textrm{id}}$ = \SI{400}{kHz} are shown in Fig. \ref{FIG5}b (Middle).
With ICG, a clear response was observed (blue line), whereas without ICG no clear response was observed (red line) due to a large noise floor at this high bandwidth.
%![Revision] S6->S8
%With ICG, the SNR of detected current response increased linearly with increasing $V_{\textrm{b}}$ (Fig. \ref{FIG5}c, blue plots; Supplementary material, SI 4, Fig. S6), although the instability of detected $I_{\textrm{i}}$ was confirmed at $V_{\textrm{b}} >$ \SI{0.5}{V} (not shown).
With ICG, the SNR of detected current response increased linearly with increasing $V_{\textrm{b}}$ (Fig. \ref{FIG5}c, blue plots; \href{https://aip.scitation.org/doi/suppl/10.1063/1.5118360/suppl_file/rsi_sup_%28sw%29191112%28clear%29.pdf}{Supplementary material}, SI 4, Fig. S8), although the instability of detected $I_{\textrm{i}}$ was confirmed at $V_{\textrm{b}} >$ \SI{0.5}{V} (not shown).
Thus, the SNR of current detection was $\sim$8 times improved by the ICG method (Fig. \ref{FIG5}c).    

The rising and falling times of the measured current changes with ICG were indistinguishable from those of the piezodriver voltage (Fig. \ref{FIG5}b, Middle and Bottom), indicating no noticeable delay ($<$ $\sim$\SI{2}{\micro s}) in the measured current response.
Note that the physically occurring (not measured) response of current change (or the rearrangement of ion distribution) must be much faster than the response of measured current changes, because the actual response is governed by the local mass transport time in the nanospace around the pipette opening.
The response time is roughly estimated to be 133 \si{ns} by adopting the diffusion time ($\tau$) required for ion transport by a distance of $2r_{\textrm{a}} =$ \SI{20}{nm}: $2r_{\textrm{a}} = \sqrt{2D\tau}$, where $D$ is the nearly identical diffusion coefficient of K$^{+}$ and Cl$^{-}$ in water ($\sim$\SI{1.5e-9}{m^2/s})~\cite{robinson2002electrolyte}.  

Contrary to our expectation, the normalized approach curve ($I_{\textrm{i}}$ vs $d$) was nearly identical between the pipettes with and without ICG (Fig. \ref{FIG5}a) although their $R_{\textrm{p}}$ values were largely different. This indicates a nearly identical $R_{\textrm{a}}/R_{\textrm{p}}$ ratio between the two cases.
This result was confirmed by FEM simulations performed by the use of various  surface charge densities of pipette and substrate in a range of 0--\SI{20}{mC/m^2} (\href{https://aip.scitation.org/doi/suppl/10.1063/1.5118360/suppl_file/rsi_sup_%28sw%29191112%28clear%29.pdf}{Supplementary material}, SI 5).
%ピペットの外の電解質のイオン伝導度がピペット内の電解質濃度に比例することを意味しますね？Fig. 3a,bと矛盾しませんか？
%! その理解で間違いないですが、矛盾はしません。濃度勾配が平行状態になると、こうなります。

In the final part of this subsection, we considered how SICM measurements with ICG would affect the membrane potential of live cells in a physiological solution.
The ICG modulates local ion concentrations around the pipette tip end, which might induce a change in the local membrane potential only when the tip is in the close vicinity to the cell surface.
However, it is difficult to perform experimental measurements of such a transient change of the local membrane potential.
Here we estimated this change for nonexcitable HeLa cells used in this study and typical excitable cells, using the Goldman-Hodgkin-Katz voltage equation~\cite{goldman1943potential,hodgkin1949effect}.
In this estimation, extracellular ion concentrations around the tip end were obtained by FEM simulations.
Full details of this analysis are described in \href{https://aip.scitation.org/doi/suppl/10.1063/1.5118360/suppl_file/rsi_sup_%28sw%29191112%28clear%29.pdf}{Supplementary material}, SI 3.
For both cell types, we found that their local membrane potentials were changed by the pipette tip with ICG to various extents depending on the value of $V_{\textrm{b}}$ (see Supplementary material SI 3, Tab. S3) .
However, for nonexcitable HeLa cell, we expect that the net contribution of an ICG-induced local membrane potential change is negligible as the tip pore size is very small.
It may not be however negligible for excitable cells, because a local membrane potential change would trigger the opening of voltage-gated sodium ion channels and thus generate action potential, which would propagate over the cell membrane.
A quantitative estimation for this possibility is beyond the scope of the present study.
Nevertheless, our FEM simulations indicate that the ICG-induced local membrane potential change can be attenuated by the use of different $V_{\textrm{b}}$ values and/or high concentration of NaCl solution instead of \SI{4}{M} KCl solution (\href{https://aip.scitation.org/doi/suppl/10.1063/1.5118360/suppl_file/rsi_sup_%28sw%29191112%28clear%29.pdf}{Supplementary material}, SI 3, Tab. S4).

\subsection{Evaluation of Improved $v_{\textrm{z}}$}

\begin{figure}[!tb]
\includegraphics{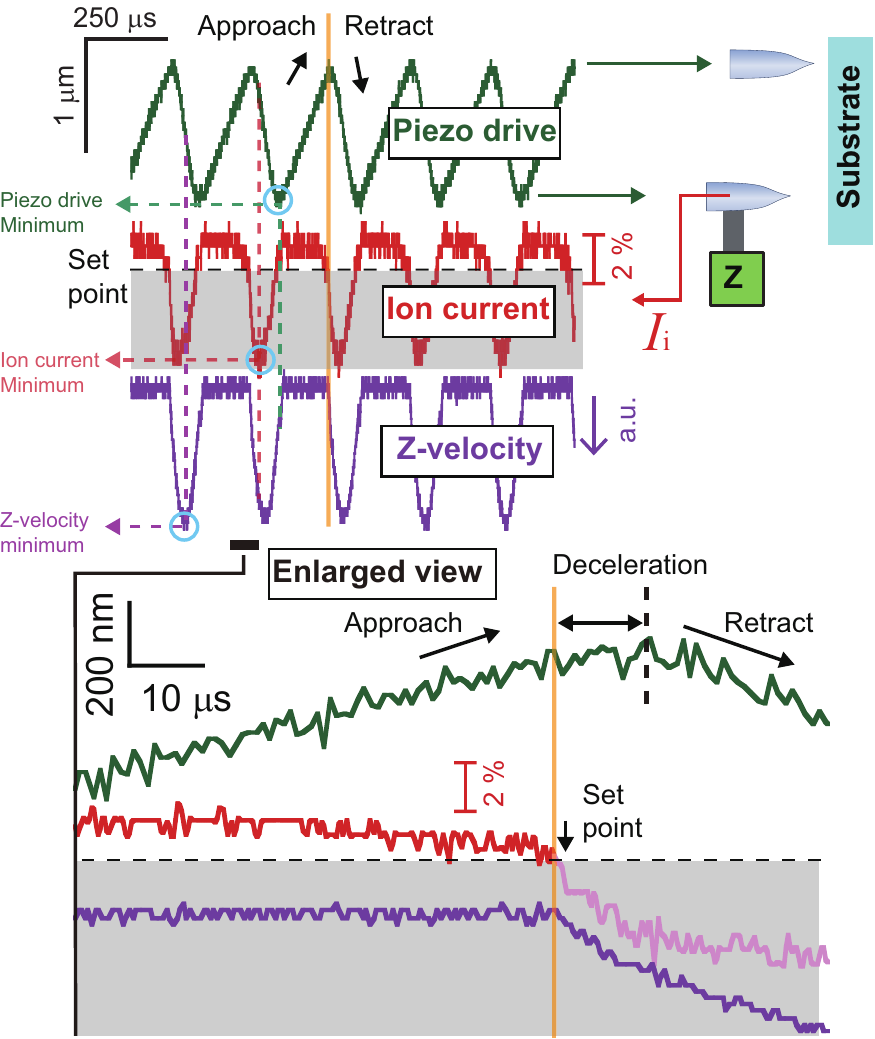}
\caption{\label{FIG6}
Evaluation of improved approach velocity. The pipette tip was periodically moved in the $z$-direction, in close proximity to the glass substrate (right panel). (Left panel) The green line indicates the time course of tip displacement estimated from the Z-scanner's drive voltage. The red line indicates the detected ion current signal. The purple line indicates the velocity of tip displacement estimated from the output current of the Z-piezodriver. (Bottom panel) An enlarged view showing these three quantities. The ion current signal in the shaded region (shown in pink) is a false one (mostly leakage current) caused by a capacitive coupling between the Z-piezoactuator and the signal line of ion current detection.
A set point value of 2$\%$ was used.
}
\end{figure}

\begin{figure*}[!ht]
    \includegraphics{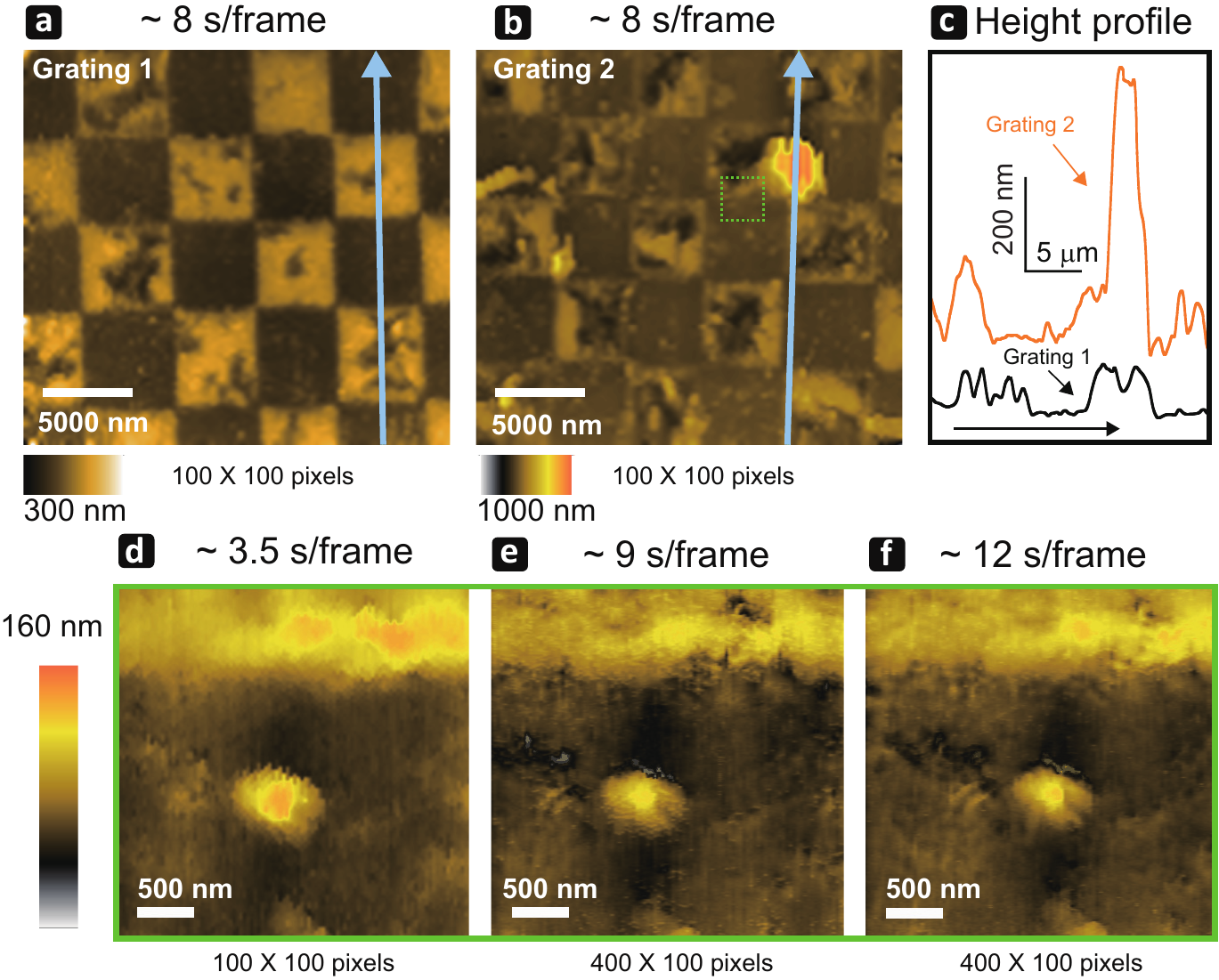}
    \caption{\label{FIG9}
    HS-SICM images captured for grating samples. (\textbf{a}, \textbf{b}) Areas of 25 $\times$ \SI{25}{\micro m^2} were imaged at $\sim$\SI{8}{s/frame} with 100 $\times$ 100 pixels for Grating 1 (\textbf{a}) and Grating 2 with a rougher surface (\textbf{b}). (\textbf{c}) Height profiles of Grating 1 and Grating 2 along the arrows shown in (\textbf{a}) and (\textbf{b}). (\textbf{d}--\textbf{f}) Images of the region shown with the small rectangle in (\textbf{b}) captured at $\sim$3.5 (\textbf{d}), $\sim$8.5 (\textbf{e}) and $\sim$\SI{12}{s/frame} (\textbf{f}).
    A set point value of 2$\%$ was used to capture these images.
    The fast scan direction is from bottom to top in these imaging experiments.}
    \end{figure*}

Here we describe a quantitative evaluation of how significantly the pipette approach velocity is improved by the ICG method and the developed Z-scanner.
Besides, we describe a problem we have encountered during this evaluation. For this evaluation, the pipette filled with \SI{4}{M} KCl was vertically moved above the glass substrate in \SI{0.15}{M} KCl solution (the Z-displacement and its velocity are shown with the green and purple lines in Fig. \ref{FIG6}, respectively), while the ion current was measured using the transimpedance amplifier with $B_{\textrm{id}}$ = \SI{400}{kHz}.
For the initiation of $\sim$\SI{1.3}{\micro m} retraction of the pipette by feedback control, the set point of ion current was set at 98$\%$ of the reference ion current (i.e., 2$\%$ reduction).
In the approaching regime, the ion current decreased as the tip got close proximity to the surface (red line in Fig. \ref{FIG6}).
%![Revision] S8->S10
%However, in the retraction regime following the deceleration phase, the detected ion current behaved strangely; $I_{\textrm{i}}$ initially decreased rather than increased and then reversed the changing direction, similar to the behavior of pipette Z-velocity (\href{https://aip.scitation.org/doi/suppl/10.1063/1.5118360/suppl_file/rsi_sup_%28sw%29191112%28clear%29.pdf}{Supplementary material}, SI 6, Fig. S8).
However, in the retraction regime following the deceleration phase, the detected ion current behaved strangely; $I_{\textrm{i}}$ initially decreased rather than increased and then reversed the changing direction, similar to the behavior of pipette Z-velocity (\href{https://aip.scitation.org/doi/suppl/10.1063/1.5118360/suppl_file/rsi_sup_%28sw%29191112%28clear%29.pdf}{Supplementary material}, SI 6, Fig. S10).
We confirmed that this abnormal response of $I_{\textrm{i}}$ was due to a leakage current caused by a capacitive coupling between the Z-piezoactuator and the signal line detecting $I_{\textrm{i}}$.
%![Revision] S8->S10
%We could mitigate this adverse effect by subtracting the gain-controlled Z-velocity signal from the measured $I_{\textrm{i}}$ (Fig. S8).
We could mitigate this adverse effect by subtracting the gain-controlled Z-velocity signal from the measured $I_{\textrm{i}}$ (Fig. S10).
Although this abnormal response could not be completely cancelled as shown with the pink line %!（色を変えてください）
%!A 対応済。
in the shaded region of Fig. \ref{FIG6}, it affected neither the feedback control nor SICM imaging. This is because the $I_{\textrm{i}}$ signal in the retraction regime is not used in the operation of SICM.      
% In the repeated approach and retraction experiments with the use of ICG method (Fig. \ref{FIG6}), we achieved $v_{\textrm{z}}$ = \SI{7.3}{\micro m/ms} for $r_{\textrm{a}}$ = \SI{25}{nm}, approximately consistent with the calculated value, 2 $\times$ \SI{25}{nm}/[\SI{1.5}{\micro s} + (\SI{400}{kHz})$^{-1}$] = \SI{9}{\micro m/ms}.
In the repeated approach and retraction experiments with the use of ICG method (Fig. \ref{FIG6}), we achieved $v_{\textrm{z}}$ = \SI{7.3}{\micro m/ms} for $r_{\textrm{a}}$ = \SI{25}{nm}, corresponding to 63$\%$ of the value estimated as 2 $\times$ \SI{25}{nm}/[\SI{1.8}{\micro s} + (\SI{400}{kHz})$^{-1}$] = \SI{11.6}{\micro m/ms}.
This $v_{\textrm{z}}$ value attained here is more than 300 times improvement over the $v_{\textrm{z}}$ value used in a recent SICM imaging study on biological samples (\SI{50}{nm/ms} for $r_{\textrm{a}}$ = \SI{50}{nm}) with a conventional design of SICM Z-scanner~\cite{novak2014imaging}. %!修正した。using ... これはbrakeboosterを使用する前の値です。
Very recently, the Sch{\"a}ffer group successfully increased $v_{\textrm{z}}$ up to \SI{4.8}{\micro m/ms} for $r_{\textrm{a}}$ = 80--\SI{100}{nm} using their sample stage scanner and a step retraction sequence called `turn step'~\cite{simeonov2019high}. Our $v_{\textrm{z}}$ value achieved for even 3--4 times smaller $r_{\textrm{a}}$ still surpasses their result. 
%! revise 191116
% We emphasize that the leakage of high salt to the outside of the tip is too small to change the ion concentrations outside and inside the tip.
We emphasize that the leakage of high salt to the outside of the tip is too small to change the bulk concentrations of ions outside and inside the tip.
% In addition, the region of ICG is confined to the vicinity of the tip for $V_{\textrm{b}}$ values ($V_{\textrm{b}} \leq |$\SI{0.1}{V}$|$) typically used in SICM measurements (Fig. \ref{FIG4}d, e), and therefore, the sample remains in the bath salt condition most of time as the time when the sample stays within a distance of $\sim$2 $\times$ $r_{\textrm{a}}$ from the tip opening is very short ($\sim$\SI{10}{\micro s}).
% The steady-state ion current of $\sim$\SI{1}{nA} %!（一桁大きくないですか？）%!A 大きくないです。
% under a bias voltage of $\sim$\SI{0.1}{V}, a typical current value used in SICM measurements, corresponds to the charge transport of $\sim$\SI{10}{fmol/s}.
%!Q（fmolは少ない印象を与えますが、tip周辺の体積は非常に小さいので、濃度に換算するととんでもなく濃くなります。もちろん、拡散しているので薄くはなりますが。Fig. 3bで薄いことを既に言っているので、この議論は不要では？ですが、$V_{\textrm{b}}$ = 0.01 Vでのシミュレーションになっているので引っかかります。何故0.5 Vでシミュレーションしないのですか？）
%!A　Supplementaryで対応する。これは電気伝導度を計測した実際の実験状況に合わせただけ。とりあえず、指示のとおりにこの議論は消去。いずれにしろ突っ込まれたら反撃できる。
In addition, the region of ICG is confined to the vicinity of the tip for $V_{\textrm{b}}$ values ($V_{\textrm{b}} \leq |$\SI{0.1}{V}$|$) typically used in SICM measurements (Figs. \ref{FIG4}d, e), and therefore, the sample remains in the bath salt condition most of time as the time when the sample stays within a distance of $\sim$2 $\times$ $r_{\textrm{a}}$ from the tip opening is very short ($\sim$\SI{10}{\micro s}).

\begin{figure*}[!tb]
    \includegraphics{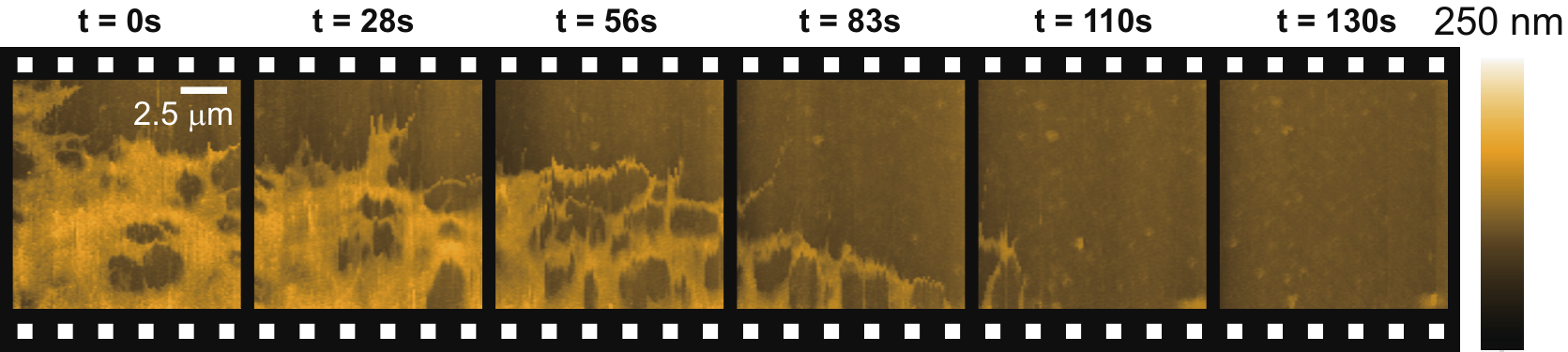}
    \caption{\label{FIG10}
    Topographic images of edge region of HeLa cell on glass substrate successively captured at 20--28 \si{s/frame} with 200 $\times$ 100 pixels, under a pixel rate of 650--\SI{1000}{Hz}, hopping amplitude of 300--\SI{500}{nm} and $V_{\textrm{b}}$ = \SI{-0.1}{V}.
    A set point value of 1.5$\%$ was used.
    The fast and slow scanning directions are from bottom to top and from left to right, respectively.
    }
\end{figure*}

\begin{figure*}[!tb]
    \includegraphics{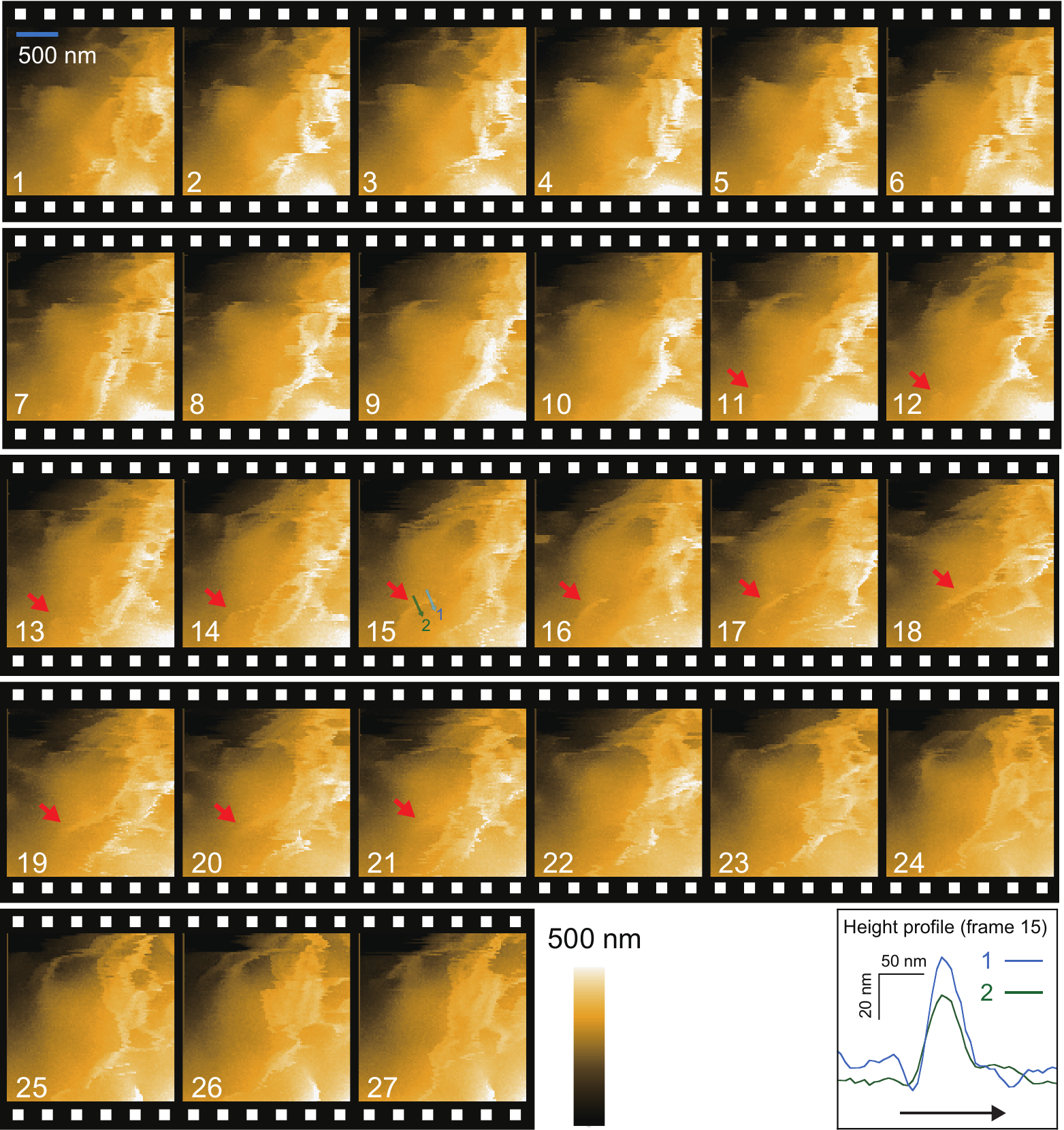}
    \caption{\label{FIG11}
    Topographic images of microvilli dynamics on HeLa cell captured with HS-SICM. These images were successively captured at $\sim$\SI{22}{s/frame} with 100 $\times$ 100 pixels, under a pixel rate of \SI{455}{Hz}, hopping amplitude of 500--\SI{600}{nm} and $V_{\textrm{b}}$ = \SI{-0.2}{V}. The numbers shown in each frame are the frame number.
    %The numbers shown in each frame are the frame number.
    The red arrows indicate a microvillus undergoing growth and disappearance. The bottom right figure indicates the height profiles along the blue (1) and green (2) arrows shown in the frame 15.
    A set point value of 1$\%$ was used to capture these HS-SICM images.
    The fast and slow scanning directions are from left to right and from bottom to top, respectively.
    }
\end{figure*}

\subsection{High-Speed Imaging of Grating Patterns}
%!Q（上でVz = 7.3 μm/msを達成したと言ったあとで、実際のイメージングではVz = 150−550 nm/msに下げてしまっているので、あれ！と感じさせますね？ra を小さくしてかなり接近させたから、とどこかで言った方がよさそうです。Bidを400 kHzから100 kHzに下げているのはどうしてですか？ノイズは400 kHzでも小さかったのでは？イメージング実験を1週間で可能ならば、ra = 20 nmくらいのピペットを使ってもっと高速なイメージングデータを出した方がよいのでは？）
%!A 20 nmで
We evaluated the performance of our HS-SICM system by capturing topographic images of a sample made of polydimethylsiloxane that had a periodic 5 $\times$ \SI{5}{\micro m^2} checkerboard pattern with a height step of \SI{100}{nm} (Grating 1).
For this imaging in hopping mode, we used $r_{\textrm{a}}$ = 5--\SI{7.5}{nm} and $B_{\textrm{id}}$ = \SI{100}{kHz}, values smaller than those used in the above evaluation test.
Therefore, we reduced $v_{\textrm{z}}$ to 150--\SI{550}{nm/ms}.
Other imaging conditions are $V_{\textrm{b}}$ = \SI{-0.3}{V} and number of pixels = 100--400 $\times$ 100.
%The imaging was carried out in hopping mode using pipettes with $r_{\textrm{a}}$ = 5--\SI{7.5}{nm} and the transimpedance amplifier with $B_{\textrm{id}}$ = \SI{100}{kHz}, under the condition of $v_{\textrm{z}}$ = 150--\SI{550}{nm/ms}, $V_{\textrm{b}}$ = \SI{-0.3}{V} and the number of pixels of 100--400 $\times$ 100.%, unless otherwise noted.
Figure \ref{FIG9}a shows a topographic image of Grating 1 captured at \SI{8}{s/frame} over a 25 $\times$ \SI{25}{\micro m^2} area with 100 $\times$ 100 pixels.
%The observed grating pitch varied by $< 10\%$ in the $X$ axis and $<$ 25$\%$ in the Y axis, arising from a hysteresis effect of the piezoactuators used for the XY scanner. It is possible to compensate for this effect by previously reported methods~\cite{watanabe2013wide,yong2009design}, we do not focus on this issue in this study.% （これはレフェリーの印象を悪くする。できるならさっさっとやればよい）
%! 画像の補正で対応する。プログラム作成したので、画像の差し替え。
Figure \ref{FIG9}b shows a topographic image of a rougher surface area of another grating sample (Grating 2) containing an object with a height of $\sim$\SI{500}{nm} (Fig. \ref{FIG9}c).
Even for this rougher surface, its imaging was possible at \SI{8}{s/frame}.
Figures \ref{FIG9}d--f show images of a narrower area of Grating 2 marked with the small rectangle in Fig. \ref{FIG9}b, captured at 3.5, 9 and \SI{12}{s/frame}, respectively.
Although fine structures were more visible in the images captured at 9 and \SI{12}{s/frame}, this difference was not due to the lower imaging rates but due to the larger number of pixels.
Averaged pixel rates were 2.85, 4.44 and  \SI{3.33}{kHz} for Figs. \ref{FIG9}d, e and f, respectively, demonstrating high temporal and spatial resolution of our HS-SICM. %!サイテーションは必要か？。
Note that the imaging rate depends not only on $v_{\textrm{z}}$ and a number of pixels but also on the hopping amplitude, hopping rate and the performance of the lateral movement of our scanner. 
In \href{https://aip.scitation.org/doi/suppl/10.1063/1.5118360/suppl_file/rsi_sup_%28sw%29191112%28clear%29.pdf}{Supplementary material}, Fig. S11, we show examples of images captured at higher rates ($\sim$4 and $\sim$\SI{0.3}{s/frame}).

\subsection{High-Speed Imaging of Biological Samples}

Next, we examined the applicability of our HS-SICM system to biological samples.
The first test sample is a live HeLa human cervical cancer cell.
The imaging was carried out in hopping mode using a pipette with $r_{\textrm{a}}$ = 5--\SI{7.5}{nm}, $B_{\textrm{id}}$ = \SI{100}{kHz} and $V_{\textrm{b}}$ = \SI{-0.1}{V}.
Figure \ref{FIG10} shows topographic images of a peripheral edge region of a HeLa cell locomoting on a glass substrate in a phosphate buffer saline, captured at 20--\SI{28}{s/frame} with 200 $\times$ 100 pixels for a scan area of 12 $\times$ 12 \si{\micro m^2}. %with 200 $\times$ 100 pixels.
During the overall locomotion downwards until the cell disappearance from the imaging area within \SI{2}{min}, the sheet-like structures (lamellipodia) with $\sim$\SI{100}{nm} height were observed to grow and retract.
In this imaging, the pixel rate was 670--\SI{1000}{Hz}, making a large contrast with the pixel rate of $\sim$\SI{70}{Hz} used in previous hopping-mode-SICM imaging of live cells without significant surface roughness~\cite{shevchuk2012alternative,novak2014imaging}.
%! [revised]191116
    Additional HS-SICM images capturing the movement of a HeLa cell in a peripheral edge region are provided in Figs. S12 and S13 (\href{https://aip.scitation.org/doi/suppl/10.1063/1.5118360/suppl_file/rsi_sup_%28sw%29191112%28clear%29.pdf}{Supplementary material}, SI 7).
    %\textcolor{red}{
        Bright-field optical microscope images before and after these SICM measurements are also provided in Fig. S14.
    %}
    In these imaging experiments, $r_{\textrm{a}}$ = 2--\SI{3}{nm}, $B_{\textrm{id}}$ = \SI{20}{kHz}, $V_{\textrm{b}}$ = \SI{70}{mV}, pixel rate = \SI{250}{Hz}, and hopping amplitude = \SI{350}{nm} are used.

%!Q (図のキャプションには0.77-1.25 kHzとありますが、 20−30 s/frameと200 × 100 pixelsの値に矛盾します。訂正してください) 
%!A (averaged) pixel rate と hopping frequencyは等しくないので、これは受け入れられない。

To demonstrate the applicability of our HS-SICM system also to live cells with significant surface roughness, we next imaged a central region (2 $\times$ \SI{2}{\micro m^2}) of a HeLa cell at \SI{22}{s/frame}, using pipettes with $r_{\textrm{a}}$ = 5--\SI{7.5}{nm}, $B_{\textrm{id}}$ = \SI{100}{kHz}, $V_{\textrm{b}}$ = \SI{-0.2}{V}, and hopping amplitude of \SI{600}{nm} (Fig. \ref{FIG11} and Supplementary Movie 1).
%!Q　キャプションにある数字は幅をもっていますが、ひとつにしました。また、Hopping frequencyはImaging rateとピクセル数から出しました。キャプションを変更してください
%!A 安藤先生のhopping frequencyの算出方法は間違いがあるため、hopping amplitudeの値を書くことにする。
The captured images show moving, growing and retracting microvilli with straight-shaped and ridge-like structures~\cite{seifert2015comparison,ida2017high,gorelik2003dynamic}.
%!Q 参考文献番号54は間違いで55ですね？白と黒の矢印を省いてください
%!A 間違いかどうか確認できませんでしたが、形状に関するものを追加しました。
The red arrows on frames 11--21 in Fig. \ref{FIG11} indicate the formation and disappearance of a single microvillus.
%During this dynamic process captured with a pixel rate of \SI{0.56}{kHz}, the full width at half maximum (FWHM) of the microvillus was less than 50 nm (lower right in Fig. \ref{FIG11}), when it was analyzed for the frame 15.
%! [Revision]
    During this dynamic process captured with a pixel rate of \SI{0.56}{kHz}, the full width at half maximum (FWHM) of the microvillus was less than \SI{50}{nm} (lower right in Fig. \ref{FIG11}), when it was analyzed for the frame 15 (FWHM values of trace 1 and 2 are 38.8 $\pm$ \SI{10.2}{nm} and 42.3 $\pm$ \SI{10.2}{nm}, respectively).
%!revision this value -> these values
These values are less than the single pixel resolution in previous SICM measurements (100--\SI{200}{nm})~\cite{seifert2015comparison,ida2017high}.
As demonstrated here, we can get SICM images with higher spatial resolution without scarifying the temporal resolution, as shown in Table \ref{tab1} (compare the values of pixel rate and $r_{\textrm{a}}$ among the three studies with respective HS-SICM systems developed).

%![Revision]
    % Note that the image distortions along the fast scan axis were found in HS-SICM images of Figures \ref{FIG10} and \ref{FIG11}.
    % They are not due to the limitation of a slow time response of the scanner used (see, Figures S3 and S11).
    % There are no noticeable distortions in the Figure \ref{FIG9}a even using higher imaging rate than that in Figures \ref{FIG10} and \ref{FIG11}. 
    % The possible explanations are some interactions with the pipette and the sample, which might be caused by fast movements of cell membranes.
    % Although, in our HS-SICM measurements, there were no significant decrease in the measured ion current that implies the collision between the pipette and cell membranes, such the interaction could modulate the SICM topography~\cite{zhou2018nanoscale,del2014contact}.
    % We speculate that such the image distortions do not frequently appear in SICM images with a large aperture pipette ($r_a$ $\sim$ \SI{100}{nm}, typically used in SICM measurements) since a small aperture pipette can visualize the tiny movements that cannot be sensed with a large aperture one.
    In Figs. \ref{FIG10} and \ref{FIG11}, slight discontinuities of images appear as noises between adjacent fast scan lines.
    In contrast, there are no such discontinuities in the images of stationary grating samples (Fig. \ref{FIG9}a) captured with an even higher imaging rate than those used in Figs. \ref{FIG10} and \ref{FIG11}.
    Moreover, during these imaging experiments, significant ion current reductions that could be caused by tip--sample contact~\cite{ida2017high} were not observed.
    Therefore, we speculate that the image discontinuities appeared in the images of live HeLa cells are due to autonomous movement of the cells and the small pipette aperture; small movements of HeLa cells that cannot be detected with a large aperture pipette ($r_a$ $\sim$ \SI{100}{nm}) appear in our high resolution images.

\begin{table*}[!htb]
    \caption{Comparison between three HS-SICM systems used for live cell imaging. The pixel rate is calculated by dividing the imaging rate by the number of pixels. $r_{\textrm{a}}$ indicates the aperture radii of the pipettes used to obtain SICM images. The attainable spatial resolution of SICM can be roughly estimated as 3 $\times$ $r_{\textrm{a}}$~\cite{rheinlaender2015lateral}. PRDA is defined as pixel rate divided by $r_{\textrm{a}}$.
    }
    \begin{tabular}{llllll}
    \hline
    Observation                                                      &\hspace{2mm} Image rate (s/frame) &\hspace{2mm} Pixel rate (s$^{-1}$) &\hspace{2mm} $r_{\textrm{a}}$ (nm) &\hspace{2mm} PRDA (s$^{-1}$nm$^{-1}$) &\hspace{2mm} Reference       \\ \hline \hline
    \begin{tabular}[c]{@{}l@{}}Endocytosis\\ Exocytosis\end{tabular} &\hspace{2mm} 6                    &\hspace{2mm} 68               &\hspace{2mm} 50                &\hspace{2mm} 1.4 &\hspace{2mm} Shevchuk et al.~\cite{shevchuk2012alternative} \\
        Peripheral edge                                                  &\hspace{2mm} 0.6               &\hspace{2mm} 1707        &\hspace{2mm} 80--100           &\hspace{2mm} 17--21 &\hspace{2mm} Simeonov and Sch{\"a}ffer~\cite{simeonov2019high}       \\
    Peripheral edge                                                  &\hspace{2mm} 20--28               &\hspace{2mm} 714--1000        &\hspace{2mm} 5--7.5            &\hspace{2mm} 95--200 &\hspace{2mm} This work       \\ \hline
    Microvilli                                                       &\hspace{2mm} 18                   &\hspace{2mm} 228              &\hspace{2mm} 50--100          &\hspace{2mm} 2.3--4.6 &\hspace{2mm} Ida et al.~\cite{ida2017high}      \\
    Microvilli                                                       &\hspace{2mm} 1.4                   &\hspace{2mm} 2926              &\hspace{2mm} 80--100          &\hspace{2mm} 29.3--36.6 &\hspace{2mm} Simeonov and Sch{\"a}ffer~\cite{simeonov2019high}      \\
    Microvilli                                                       &\hspace{2mm} 20                   &\hspace{2mm} 455              &\hspace{2mm} 5--7.5             &\hspace{2mm} 60.7--91.0 &\hspace{2mm} This work       \\ \hline
    \end{tabular}
    \label{tab1}
\end{table*}

\begin{figure*}[!htb]
    \includegraphics{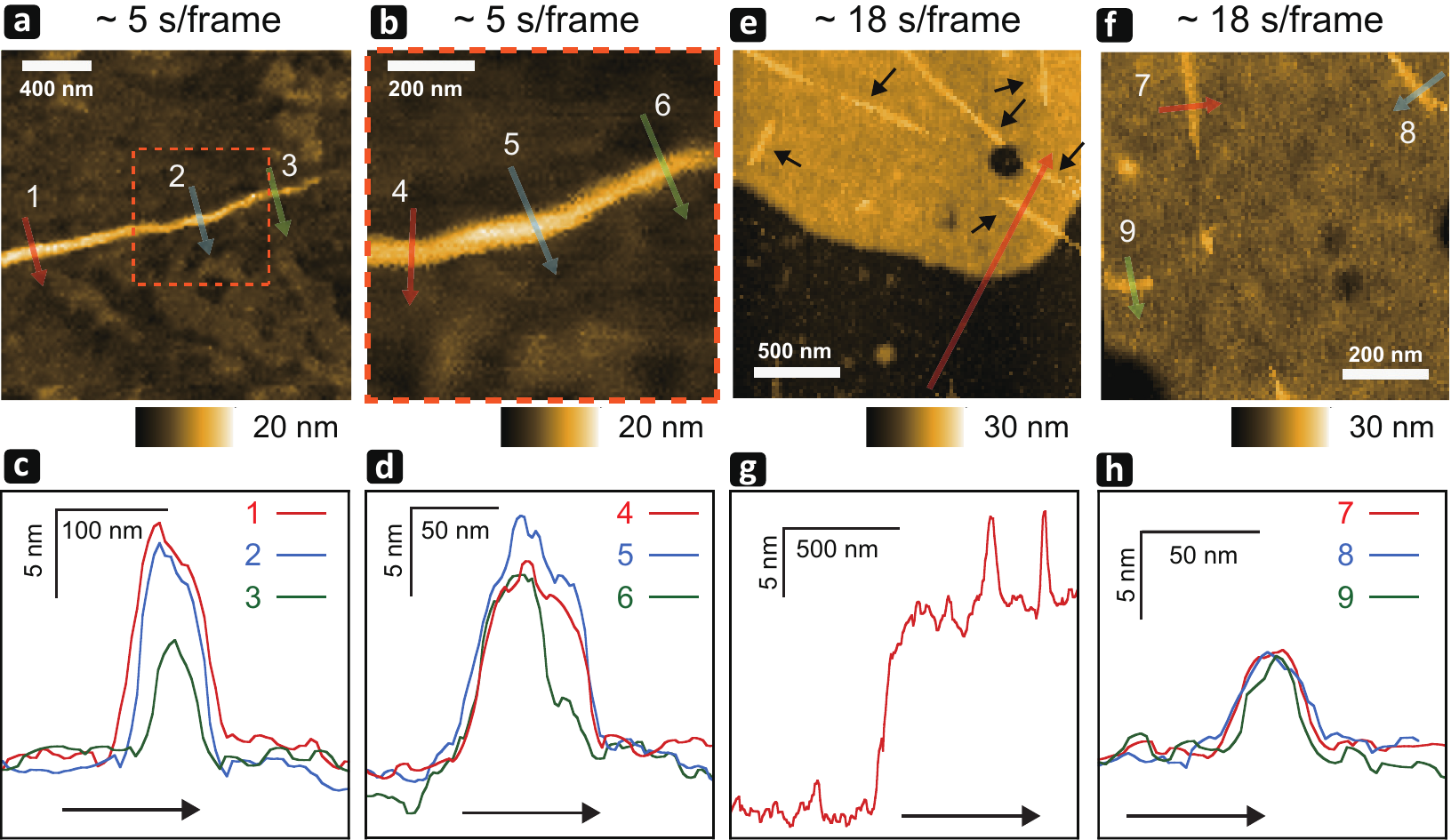}
    \caption{
    HS-SICM images of actin filaments. (\textbf{a}--\textbf{d}) Actin filaments were attached onto a glass surface coated with positively charged amino-propyltriethoxysilane. (\textbf{e}--\textbf{h}) Partially biotinylated actin filaments were attached onto a mica-supported lipid bilayer containing a biotin-lipid through streptavidin molecules with a low surface density. (\textbf{a}) HS-SICM image captured at $\sim$\SI{5}{s/frame} with 100 $\times$ 100 pixels, a pixel rate of \SI{2000}{Hz}, hopping amplitude of \SI{100}{nm} and $V_{\textrm{b}}$ = \SI{-0.2}{V}. (\textbf{b}) A magnified view of the rectangle area shown in (\textbf{a}). (\textbf{c}) Height profiles along the arrows shown in (\textbf{a}). (\textbf{d}) Height profiles along the arrows shown in (\textbf{b}). (\textbf{e}, \textbf{f}) HS-SICM images captured at $\sim$\SI{18}{s/frame} with 100 $\times$ 100 pixels, a pixel rate of \SI{556}{Hz}, hopping amplitude of \SI{250}{nm} and $V_{\textrm{b}}$ = \SI{-0.2}{V}. (\textbf{g}) Height profile along the arrow shown in (\textbf{e}). (\textbf{h}) Height profiles along the arrows shown in (\textbf{f}).
    A set point value of 1.5$\%$ was used to capture these HS-SICM images.}
    \label{FIG8}
\end{figure*}
%! xxを埋めるべし。

Next, we performed HS-SICM imaging of actin filaments of $\sim$\SI{7}{nm} in diameter, under the conditions of $r_{\textrm{a}}$ = 5--\SI{7.5}{nm}, $B_{\textrm{id}}$ = \SI{100}{kHz}, $V_{\textrm{b}}$ = \SI{-0.2}{V}, pixel rate of 556--\SI{2000}{Hz}, and hopping amplitude of 100--\SI{250}{nm}.
Figure \ref{FIG8}a shows a topographic image captured at $\sim$\SI{5}{s/frame} of an actin filament specimen placed on a glass substrate coated with positively charged aminopropyl-triethoxysilane.
Figure \ref{FIG8}b shows its enlarged image for the area shown with the red rectangle in Fig. \ref{FIG8}a.
The image exhibited a height variation along the filament, as indicated in Figs. \ref{FIG8}c and d.
The measured height obtained from the arrow position 3 is $\sim$\SI{7}{nm}.
%This value corresponds to the height of the single actin filament. 
However, the measured heights obtained from the arrow positions 1, 2, 4, 5, and 6 were around \SI{14}{nm}.
These results may indicate that the specimen partly contains vertically stacked two actin filaments.
%The measured FWHM was $\sim$\SI{35}{nm} for the arrow position 3.
%! [Revision]
The measured FWHM was 38.1 $\pm$ \SI{4.2}{nm} for the arrow position 3.
This value is 5 times larger than the diameter of an actin filament. 
This result can be explained by the side wall effect~\cite{dorwling2018simultaneous}; the tip wall thickness, i.e., $r_{\textrm{o}}-r_{\textrm{i}}$, would expand the diameter of a small object measured with SICM~\cite{rheinlaender2015lateral}.
%The side wall effect also explains that the measured FWHMs for the arrow positions 1 ($\sim$\SI{60}{nm}) and 2 ($\sim$\SI{55}{nm}) were larger than that for the arrow position 3.
%! [Revision]
The side wall effect also explains that the measured FWHMs for the arrow positions 1 (66.0 $\pm$ \SI{2.6}{nm}) and 2 (50.6 $\pm$ \SI{3.7}{nm}) were larger than that for the arrow position 3.
% The measured FWHM was $\sim$\SI{60}{nm}, $\sim$\SI{55}{nm} and $\sim$\SI{35}{nm} for the arrow position 1, 2 and 3, respectively.
%!Q （何故position 3でのFWHMを示さないのでしょうか？position 3ではsingle filamentの可能性が高いので、XY方向の空間分解能を議論するのに適していますよね？）
%!A すみません。間違いました。これはarrow position 3のことです。
% this value is 5 times larger than the diameter of the actin filament. 
% This result can be explained by the side wall effect~\cite{dorwling2018simultaneous}; the tip wall thickness, i.e., $r_{\textrm{o}}$ - $r_{\textrm{i}}$, would expand the diameter of a small objects measured with SICM~\cite{rheinlaender2015lateral}.
Despite the pixel size of 8 $\times$ \SI{8}{nm^2}, the \SI{36}{nm} crossover repeat of the two-stranded actin helix could not be resolved.
This is not due to insufficient vertical resolution but due to insufficient lateral resolution of the pipette used.

Next, we used mica-supported neutral lipid bilayers containing biotin-lipid, instead of using the amino silane-coated glass substrate to avoid possible bundling of actin filaments on the positively charged surface.
Figures \ref{FIG8}e, f show images captured at \SI{18}{s/frame} for partially biotinylated actin filaments immobilized on the lipid bilayers through streptavidin molecules with a low surface density.
Measured heights of these filaments were $\sim$6--\SI{7}{nm} (Figs. \ref{FIG8}g and h).
However, the measured height of the lipid bilayer was $\sim$\SI{13}{nm} from the mica surface (Fig. \ref{FIG8}), much larger than the bilayer thickness of $\sim$\SI{5}{nm}~\cite{leonenko2004investigation}.
This large measured thickness is possibly due to a sensitivity of $I_{\textrm{i}}$ to negative charges on the mica surface~\cite{mckelvey2014surface,klausen2016mapping,perry2016surface,fuhs2018direct}.
The surface charges of objects can change $d_{\textrm{c}}$ even at constant  $V_{\textrm{b}}$ and constant set point~\cite{klausen2016mapping}, which can provide measured height largely deviating from real one.
%and therefore, the apparent height measured by SICM changes.
%In the high resolution image of immobilized filaments (Fig. \ref{FIG8}f), measured values for height and FWHM were $\sim$\SI{6}{nm} and 25--\SI{30}{nm}, respectively (Fig. \ref{FIG8}h).
%![Revision]
In the high resolution image of immobilized filaments (Fig. \ref{FIG8}f), the measured value of FWHM was 33.3 $\pm$ \SI{6.0}{nm} (Fig. \ref{FIG8}h).

As demonstrated here, our HS-SICM enables fast imaging of molecules without sacrificing the pixel resolution, unlike previous works~\cite{novak2014imaging,shevchuk2012alternative}.
%When the number of pixels were reduced to 50 $\times$ 50 for a 0.8 $\times$ \SI{0.8}{\micro m^2} area, we could image a low-height sample at $\sim$\SI{0.9}{s/frame} (\href{https://aip.scitation.org/doi/suppl/10.1063/1.5118360/suppl_file/rsi_sup_%28sw%29191112%28clear%29.pdf}{Supplementary material}, SI 8 and Supplementary Movie 2).  
%! [Revision]　説明追加
    % When the number of pixels were reduced to 50 $\times$ 50, we could get sub-second image rate for a low-height sample.
    % In Supplementary Movie 2, small polymers formed from the silane coupling agent on mica substrate were captured for a 0.8 $\times$ \SI{0.8}{\micro m^2} area.
    % HS-SICM could capture noticeable morphological changes of the polymer samples caused by a high fluidity of the polymer sample.
    When the number of pixels was reduced to 50 $\times$ 50, we could achieve sub-second imaging for a low-height sample.
    \href{https://aip.scitation.org/doi/suppl/10.1063/1.5118360/suppl_file/movie_2.mp4}{Supplementary Movie 2} captured at \SI{0.9}{s/frame} with 50 $\times$ 50 pixels over a 0.8 $\times$ \SI{0.8}{\micro m^2} area shows high fluidity-driven morphological changes of polymers formed from a silane coupling agent placed on mica.

\subsection{Outlook for Higher Spatiotemporal Resolution}

Finally, we discuss further possible improvements of HS-SICM towards higher spatiotemporal resolution.
The speed performance of SICM can be represented by the value of pixel rate divided by $r_{\textrm{a}}$ (we abbreviate this quantity as PRDA), because of a trade-off relationship between temporal resolution and spatial resolution.
In Table \ref{tab1}, PRDA values of our HS-SICM imaging are shown, together with those of HS-SICM imaging in other labs.
Although PRDA depends on sample height, our HS-SICM system set a highest record, PRDA = 95--\SI{200}{Hz/nm}, in the imaging of a peripheral edge region of a HeLa cell.
This record was attained by two means: the ICG method granting the high SNR of current detection and the high resonance frequency of the Z-scanner.
Since we have not yet introduced other devices proposed previously for increasing the temporal resolution, there are still room for further speed enhancement.
One of candidates to be added is (i) the `turn step' procedure (applying a step function to the Z-piezodriver) developed by Simeonov and Sch{\"a}ffer for rapid pipette retraction~\cite{simeonov2019high}.
Other candidates would be (ii) further current noise reduction of the transimpedance amplifier in a high frequency region, and (iii) lock-in detection of AC current produced by modulation of the pipette Z-position with small amplitude~\cite{pastre2001characterization}. Since our Z-scanner has a much higher resonance frequency than ever before, we will be able to use high-frequency modulation ($\sim$\SI{100}{kHz}) to achieve faster lock-in detection of AC current.
For higher spatial resolution, we need to explore methods to fabricate a pipette with smaller $r_{\textrm{a}}$ and $r_{\textrm{o}}$, without significantly increasing $R_{\textrm{p}}$.
One of possibilities would be the use of a short carbon nanotube inserted to the nanopore of a glass pipette with low $R_{\textrm{p}}$.

\section{Conclusions}

HS-SICM has been desired to be established not only to improve the time efficiency of imaging but also to make it possible to visualize dynamic biological processes occurring in very soft, fragile or suspended (not on a substrate) samples that cannot be imaged with HS-AFM.
As demonstrated in this study, the fast imaging capability of SICM can be achieved by the improvement of speed performances of pipette Z-positioning and ion current detection.
The former was attained by the new Z-scanner and implementation of vibration damping techniques to the Z-scanner.
The latter was attained by the minimization of the total capacitance at the amplifier input and by the reduction of $R_{\textrm{p}}$ achieved with the ICG method, resulting in an increased SNR of ion current detection.
The resulting $v_{\textrm{z}}$ reached \SI{0.55}{\micro m/ms} for $r_{\textrm{a}}$ = 5--\SI{7.5}{nm} and \SI{7.3}{\micro m/ms} for $r_{\textrm{a}}$ = \SI{25}{nm}.
The value of \SI{7.3}{\micro m/ms} is larger than the recent fastest record achieved by Simeonov and Sch{\"a}ffer: \SI{4.8}{\micro m/ms} for $r_{\textrm{a}}$ = 80--\SI{100}{nm}.
Consequently, the highest possible imaging rate was enhanced by $\sim$100-times, compared to conventional SICM systems.
Even sub-second imaging is now possible for a scan area of 0.8 $\times$ \SI{0.8}{\micro m^2} with 50 $\times$ 50 pixels, without compromise of spatial resolution.
The achieved speed performance will contribute to the significant extension of SICM application in biological studies.

\section{Methods}

\subsection{Fabrication of Nanopipettes}

%! version 190128
%Nanopipettes were pulled from quartz capillaries QF100-70-7.5 (\SI{1.0}{mm} o.d., \SI{0.50}{mm} i.d. with filament) and Q100-30-15 (\SI{1.0}{mm} o.d., \SI{0.30}{mm} i.d.) (both Sutter Instrument) using a P-2000 laser puller (Sutter Instrument). Just before pulling, each capillary was softly plasma-etched using a plasma etcher (SOUTH BAY TECHNOLOGY) at \SI{20}{W} under an oxygen gas flow at a pressure of \SI{120}{mTorr} for \SI{5}{min}. The size and cone angle of the pipette tips were estimated using a scanning electron microscope (ZEISS) and electrical resistance measurements. After backfilling, the nanopipette from the quartz capillary of Q100-30-15 was subjected to a pressure of \SI{0.2}{MPa} for \SI{60}{min} to obtain tip sufficient conductance. The nanopipettes from QF100-70-7.5 were used for measuring the tip conductance in Fig. \ref{FIG4}a.

We prepared pipettes by pulling laser-heated quartz glass capillaries, QF100-70-7.5 (outer diameter, \SI{1.0}{mm}; inner diameter, \SI{0.50}{mm}; with filament) and Q100-30-15 (outer diameter, 1.0 \si{mm}; inner diameter, \SI{0.30}{mm}; without filament) from Sutter Instrument, using a laser puller (Sutter Instrument, P-2000). Just before pulling, we softly plasma-etched for \SI{5}{min} at \SI{20}{W} under oxygen gas flow (\SI{120}{mTorr}), using a plasma etcher (South Bay Technology, PE2000) to remove unwanted contamination inside the pipette.
%The size and cone angle of each pipette tip were estimated from its scanning electron micrograph (Zeiss, SUPRA 40VP) and measured electrical resistance.
%![Revision]
The size and cone angle of each pipette tip were estimated from its scanning electron micrographs (Zeiss, SUPRA 40VP), transmission electron micrographs (JOEL, JEM-2000EX), and measured electrical resistance.
Pipettes prepared from QF100-70-7.5 were used for the conductance measurement shown in Fig. \ref{FIG4}c.

%\subsection{Measurement of Transfer Function and Time Domain Response}
\subsection{Measurements of Z-scanner Transfer Function and Time Domain Response}

%The displacement of the nanopositioner was measured with a laser vibrometer (NLV-2500, Polytech or ST-3761, IWATSU). To obtain the transfer function of the Z-nanopositioner, we used a combination of an E5106B network analyzer (Agilent Technology) and a laser vibrometer. A square-like-waveform voltage generated with a WF1948 function generator (NF) was used for measuring the time domain response. The output waveform of the function generator was amplified with an M-2141 piezodriver (gain; $\times$ 15, bandwidth; 1 MHz, MESTEK).

The Z-scanner displacement was measured with a laser vibrometer (Polytech, NLV-2500 or Iwatsu, ST-3761). The transfer function characterizing the Z-scanner response was obtained using an network analyzer (Agilent Technology, E5106B). A square-like-waveform voltage generated with a function generator (NF Corp., WF1948) and then amplified with a piezodriver (MESTECK, M-2141; gain, $\times$15; bandwidth, \SI{1}{MHz}) was used for the measurement of time domain response of the Z-scanner.
%! Q:(pipetteのdisplacementは測定していない？) 
%! A: 測定してないです。標準試料で高さの構成はしています。

%\subsection{Measurement of Conductance of the Tip under Ion Concentration Gradient}
\subsection{Measurement of Pipette Conductance with and without ICG}

%To evaluate the enhancement of the tip conductance $\rho ^{-1}$ due to ICG, we prepared a set of two pipette tips fabricated from a single capillary tube using the laser pulling apparatus. Although the nanopore diameter of the tip is inevitably varied between fabrication runs, the pair of tips produced in a single fabrication run have nearly identical nanopore diameter within $\pm$10$\%$, which is confirmed by electrical measurements. We compared the conductivity between the one pipette of the pair with ICG and the other pipette without ICG. The enhanced factor is obtained from the ratio of conductances. To obtain the single point in Fig. \ref{FIG4}c, we measured tip conductances produced at different fabrication runs at least n $>$ 5. The value of the conductance was obtained in a range of the bias application of \SI{-10}{mV} $\leq V_{\textrm{b}} \leq$ \SI{10}{mV} to avoid the problem of a non-linear potential--current relationship originating from the ion current rectification effect. These procedures help to determine the conductance enhancement due to ICG.
To evaluate the enhancement of pipette conductance (1/$R_{\textrm{p}}$) by ICG, we prepared a pair of pipettes simultaneously produced from one pulled capillary, which exhibited a difference in $r_{\textrm{a}}$ less than $\pm$10$\%$, as confirmed by electrical conductance measurements under an identical condition.
One of the pair of pipettes was applied to a conductance measurement under ICG, while the other to a conductance measurement without ICG. To obtain each plot in Fig. \ref{FIG4}c, we measured (1/$R_{\textrm{p}}$) for more than 5 sets of pipettes.
%! revision between xx to xx. -> xx and xx
To avoid the non-linear current-potential problem arising from an ion current rectification effect, we used $V_{\textrm{b}}$ ranging between \SI{-10}{mV} and \SI{10}{mV}.

%\subsection{Measurement of Approach Curves and SNR Induced by Vertical Tip Movement}
\subsection{Measurements of Approach Curves and Response of Ion Current}

%To obtain the approach curves shown in Fig. \ref{FIG5}a, a digital low-pass filter (6th order, cut-off frequency of \SI{10}{kHz}) was used. After the curve was traced with $V_{\textrm{b}}$ = \SI{0.1}{V}, the tip was moved at the point that shows 5$\%$ reduction of the unmodulated ion current calculated from the approach curve just taken. Then, we shifted the cut-off frequency of the low-pass filter to \SI{400}{kHz} to capture a fast response of ion current changes and $V_{\textrm{b}}$ was set to measurement values. After this procedure was performed, we started the experiments shown as Fig. \ref{FIG5}b. In addition, an identical tip was used to avoid any influence of the tip shape between experiments. The experiment without ICG was carried out first. Then, the solution inside the tip was replaced by a high-salt solution. After the tip was immersed in a high-salt solution for a sufficiently long time, then the tip was immersed in a physiological salt concentration for a sufficiently long time. After that, we carried out the experiment with ICG.

To obtain the approach curves ($I_{\textrm{i}}$ vs $d$) shown in Fig. \ref{FIG5}a, a digital 6th-order low-pass filter was used with a cutoff frequency of \SI{10}{kHz}. After the measurement of each curve under $V_{\textrm{b}}$ = \SI{0.1}{V}, the pipette was moved to a Z-position where the ion current reduction by 5$\%$ had been detected in the approach curve just obtained. Next, the cut-off frequency of the low-pass filter was increased to 400 kHz for the measurement of fast ion current response and  $V_{\textrm{b}}$  was set at a measurement value. Then, the experiment shown in Fig. \ref{FIG5}b was performed. This series of measurements were repeated under different values of  $V_{\textrm{b}}$  and with/without ICG. The identical pipette was used through the experiments to remove variations that would arise from varied pipette shapes. After completing the experiments without ICG, the KCl solution inside the pipette was replaced with a \SI{4}{M} KCl solution by being immersed the pipette in a \SI{4}{M} KCl solution for a sufficiently long time ($>$ \SI{60}{min}). The $R_{\textrm{p}}$ value of the pipetted with ICG prepared in this way was confirmed to be nearly identical to that of a similar pipette filled with a \SI{4}{M} KCl solution from the beginning. 
%! Q：(~XX hrs 本当に置換できるのでしょうか？以下の確認はしましたか？)
%! A：できます。熱拡散で考えても十分妥当です。実際測定しても30分くらい待てば電気伝導が落ち着きます。

\subsection{HS-SICM Apparatus}

The HS-SICM apparatus used in this study was controlled with home-written software built with Labview 2015 (National Instruments), which was also used for data acquisition and analysis. The HS-SICM imaging head includes the XYZ-scanner composed of AE0505D08D-H0F and AE0505D08DF piezoactuators for the Z and XY directions (both NEC/tokin), respectively,  as shown in Fig. \ref{FIG2}a. The Z- and XY-piezoactuators were driven using M-2141 and M-26110-2-K piezodrivers (both MESTEK), respectively. The overall control of the imaging head was performed with a home-written FPGA-based system (NI-5782 and NI-5781 with NI PXI-7954R for the Z- and XY-position control, respectively; all National Instruments). For coarse Z-positioning, the imaging head was vertically moved with an MTS25-Z8 stepping-motor-based linear stage (travel range, \SI{25}{mm}; THORLABS). %For the FB and FF control and noise filtering, homemade analog and FPGA-integrated circuits were used.
For the FF control, FPGA-integrated circuits were used, while homemade analog circuities were used for FB and noise filtering.
The sample was placed onto the home-built XY-coarse positioner with a travel range of \SI{20}{mm}, which was placed onto an ECLIPSE Ti-U inverted optical microscope (Nikon). The ion current through the tip nanopore was detected via transimpedance amplifiers CA656F2 (bandwidth, \SI{100}{kHz}; NF) and LCA-400K-10M (bandwidth, \SI{400}{kHz}; FEMTO). A WF1948 function generator (NF) was used for the application of tip bias potential.

\subsection{HS-SICM Imaging in Hopping Mode}

The tip was approached to the surface with $v_{\textrm{z}}$ (0.05--\SI{7}{\micro m/ms}) until the ion current reached a set point value.
%The set point was typically set at 1$\%$ of the “reference current” flowing when the tip was well far away from the surface (10--\SI{100}{pA} when the ICG method was adopted).
%![Revision]
The set point was set at 1--2$\%$ reduction of the ``reference current'' flowing when the tip was well far away from the surface (10--\SI{100}{pA} when the ICG method was adopted).
% \textcolor{red}{
% The setpoint values are 2, 2, 1.5, 1, 1.5, 2, 5, 5, and 2$\%$ in Figures \ref{FIG6}, \ref{FIG9}, \ref{FIG10}, \ref{FIG11}, \ref{FIG8}, S11, S12, S13 and S15, respectively.
% }
The voltage applied to the Z-scanner yielding the set point current was recorded as the sample height at the corresponding pixel position. Here, the output from the transimpedance amplifier was high-pass filtered at 5--\SI{1000}{Hz} to suppress the effect of current drift on SICM imaging. Then, the pipette was retracted by a hopping distance (\SI{20}{nm}--\SI{50}{\micro m} within 20--\SI{50}{\micro s}, depending on the hopping amplitude, during which the pipette was moved laterally towards the next pixel position. After full retraction, the tip approaching was performed again.
%![Revision]
Through all the HS-SICM experiments, the pipette resistance did not show a significant change, indicating no break of the pipette tip during scanning.
%In some cases, the synchronization between the Z- and XY displacements were removed, meaning that nearest pixels along the fast-scan axis are sometimes identical owing to insufficient $v_{\textrm{z}}$（意味が分かりません）. In this mode, the tip is always vertically oscillated with constant amplitude. This operation is useful for suppressing the effect of the long-term drift of the Z-scanner on the imaging.
%! Q:(hopping frequencyとは無関係？)
%! A:時として問題になるので、調整しています。

%! Q:（意味が分かりません）
%! A:決定的な必要性を感じないので消します。

%To capture topographic images in this study, we used the following setup, unless otherwise noted: nanopore diameter $\sim$ 10--20 nm, bias potential $\sim-0.1$--$-0.3$ V, bandwidth of ion current detector $\sim$ \SI{100}{kHz}, fall velocity $\sim$ 200--500 nm/ms, and hopping amplitude $\sim$ 20--100 nm.
%! Q:この部分はそれぞれのFigureのLegendに書くべき。
%! A:そのとおりだと思いますのでそうします。

%![Revision]
%\subsection{Image Analysis}
The value of FWHM was calculated from five height profiles of each HS-SICM image.
The error of FWHM was estimated from the standard deviation.

\subsection{Sample Preparation}

\noindent \textbf{(1) Glass substrate}\\
%Cover slips (C024321, Matsunami Glass) were used as glass substrate in this study. \\
Cover slips (Matsunami Glass, C024321) cleaned with a piranha solution for 60 min at \SI{70}{\celsius} were used as a glass substrate.\\
%! Q： 購入したものを洗浄せずに使ったのか？
%! A: アクチン繊維をガラス上で見たものに関しては、ピラニア洗浄済のものを使用。

\noindent \textbf{(2) HeLa cells on glass}\\
%HeLa cells were cultured in DMEM medium (Dulbecco's Modified Eagle's Medium, Gibco) supplemented with 10$\%$ fetal bovine serum. Cells were maintained in a humidified 5$\%$ CO$^2$ incubator at \SI{37}{\celsius} on MAS-coated glass (S9441, Matsunami Glass).
%HeLa cells were cultured in Dulbecco's Modified Eagle's Medium (Gibco) supplemented with 10$\%$ fetal bovine serum. The cells were deposited on a MAS-coated glass (Matsunami Glass, S9441) and maintained in a humidified 5% CO2 incubator at 37 °C until observation.
HeLa cells were cultured in Dulbecco's Modified Eagle's Medium (Gibco) supplemented with 10$\%$ fetal bovine serum. The cells were deposited on a MAS-coated glass (Matsunami Glass, S9441) and maintained in a humidified 5$\%$ CO$^2$ incubator at \SI{37}{\celsius} until observation. Then, the culture medium was changed to phosphate buffer saline (Gibco, PBS). Then, HS-SICM measurements were performed at room temperature.\\
%! Q：（SICM観察中も5%CO2, 37°Cに保てるのか？）
%! A:保てません。

\noindent \textbf{(3) HeLa cells on plastic dish}\\
HeLa cells were seeded on plastic dishes (AS ONE, 1-8549-01) in Dulbecco's Modified Eagle's Medium (Gibco) supplemented with 10$\%$ fetal bovine serum. The cells were incubated at \SI{37}{\celsius} with 5$\%$ CO$^2$ and measured by HS-SICM 3--4 days after seeding. Before HS-SICM measurements, the culture medium was changed to phosphate buffer saline (Gibco, PBS). HS-SICM measurements were performed at room temperature.\\

\noindent \textbf{(4) Actin filaments on glass substrate}\\
%The glass surface was first coated with amino silane. Then, actin filaments with a buffer solution introduced to the glass surface. The buffer solution contains \SI{25}{mM}-KCl, \SI{2}{mM} MgCl$_2$, \SI{1}{mM} EGTA, \SI{20}{mM} imidazole-HCl (pH7.6).\\
The glass surface was first coated with (3-aminopropyl) triethoxysilane (APTES; Sigma Aldrich). Then, a drop (12 \si{\micro L}) of actin filaments prepared according to the method~\cite{sakamoto2000direct} and diluted to 3 \si{\micro M} in Buffer A containing \SI{25}{mM} KCl, \SI{2}{mM} MgCl$_2$, \SI{1}{mM} EGTA, \SI{20}{mM} imidazole-HCl (pH7.6) was deposited to the glass surface and incubated for 15 \si{min}. Unattached actin filaments were washed out with Buffer A.\\

\noindent \textbf{(5) Actin filaments on lipid bilayer}\\
%Mica surface was coated with the lipid that contains 1,2-dipalmitoyl-sn-glycero-3-phosphocholine (DPPC), 1,2-dipalmitoyl-sn-glycero-3-phosphoethanolamine-N-(cap biotinyl)(biotin-cap-DPPE) in a weight ratio of 0.99:0.01. Partially biotinylated actin filaments were immobilized on the lipid bilayer surface through streptavidin with a low surface density. The buffer solution in actin filaments is same one mentioned the above.
The mica surface was coated with lipids containing 1,2-dipalmitoyl-sn-glycero-3-phosphocholine (DPPC) and 1,2-dipalmitoyl-sn-glycero-3-phosphoethanolamine-N-(cap biotinyl) (biotin-cap-DPPE) in a weight ratio of 0.99:0.01, according to the method~\cite{yamamoto2010high}. Partially biotinylated actin filaments in Buffer A prepared according to the method~\cite{kodera2010video} were immobilized on the lipid bilayer surface through streptavidin with a low surface density. Unattached actin filaments were washed out with Buffer A.\\

%\subsection{Finite Element Simulations}
\subsection{FEM Simulations}

We employed three-dimensional FEM simulations to study electrostatics and ionic mass transport processes in the pipette tip with ICG. In the simulation, we used the rotational symmetry along the pipette axis to reduce the simulation time. The full detail is described in SI 3. Briefly, the following set of equations were solved numerically:

%We employed three-dimensional FEM simulations to study electrostatics and ionic mass transport processes in the pipette tip under an ion ICG. In the simulation, we used the rotational symmetry along the pipette axis to reduce time to obtain the results. The full detail is described in the supplementary information. Briefly, the following sets of equations were solved numerically:

\begin{eqnarray}
\label{eq_P}
&&\nabla^2 V = - \frac{F}{\varepsilon_0 \varepsilon} \Sigma_{j =1} ^2 Z_j c_j  \\
\label{eq_NP}
&&{\bf J_j} = -D_j(c) \nabla c_j - \frac{F Z_j c_j D_j(c) }{RT} \nabla V , \nonumber \\
&&\nabla \cdot {\bf J}_j = 0 \\
\label{eq_SCD}
&&{\bf n} \cdot \nabla \phi = \frac{- \sigma}{\varepsilon_0 \varepsilon}.
\end{eqnarray}

%The Poisson equation [Eq. (\ref{eq_P})] describes the electrostatic potential $V$ and electric field with the spatial charge distribution in a continuous medium of permittivity $\varepsilon$ containing the ions $j$ of concentration $c_j$ and charge $Z_j$. $F$ and $\varepsilon_0$ are the Faraday constant and the vacuum permittivity, respectively. We assume that the movement of the tip is sufficiently slow to neglect convection; and thus obtained the time-independent Nernst--Planck equation as Eq. (\ref{eq_NP}). This equation describes the diffusion and migration of the ions, in which ${\bf J_j}$, $D_j (c)$, $R$ and $T$ are the ion flux concentration of $j$, concentration-dependent diffusion constant of $j$, gas constant, and temperature, respectively. The boundary condition for Eq. (\ref{eq_NP}) are determined as a zero flux or constant concentration condition is satisfied. On the other hand, the boundary conditions of Eq. (\ref{eq_P}) are given as a fixed potential or the spatial distribution of the surface charge $\sigma$ as [Eq. \ref{eq_SCD}], where ${\bf n}$ represents the surface normal vector.
The Poisson equation Eq. (\ref{eq_P}) describes the electrostatic potential $V$ and electric field with a spatial charge distribution in a continuous medium of permittivity $\varepsilon$ containing the ions $j$ of concentration $c_j$ and charge $Z_j$. $F$ and $\varepsilon_0$ are the Faraday constant and the vacuum permittivity, respectively. We assume that the movement of the tip is sufficiently slow not to agitate the solution, and thus, the time-independent Nernst--Planck equation, Eq. (\ref{eq_NP}), holds, where $\textbf{J}_j$, $D_j (c)$, $R$ and $T$ are the ion flux concentration of $j$, concentration-dependent diffusion constant of $j$, gas constant, and temperature in kelvin, respectively. This equation describes the diffusion and migration of the ions. The boundary condition for Eq. (\ref{eq_NP}) is determined so that a zero flux or constant concentration condition is satisfied. On the other hand, the boundary conditions of Eq. (\ref{eq_P}) are given so that a fixed potential or the spatial distribution of the surface charge $\sigma$, as described in Eq. (\ref{eq_SCD}), holds. In Eq. (\ref{eq_SCD}), $\textbf{n}$ represents the surface normal vector.
%! 式中のSCDは不要では? 入れるならどこかで定義
%! たしかに不要のように思えますね。

\section*{Supplementary Materials}

The following data are available as \href{https://aip.scitation.org/doi/suppl/10.1063/1.5118360}{Supplementary material}: performance of XYZ-Scanner, current noise in our SICM system, finite-element simulation, dynamic response of measured ion current with and without use of ICG method, simulated approach curves obtained with and without the use of ICG method, current noise caused by capacitive couplings between Z-scanner and signal line of current detection, high-speed SICM imagings of grating samples and peripheral edge of HeLa cells, HS-SICM images of microvilli dynamics of HeLa cell (\href{https://aip.scitation.org/doi/suppl/10.1063/1.5118360/suppl_file/movie_2.mp4}{Movie 1}), and HS-SICM images of polymers (\href{https://aip.scitation.org/doi/suppl/10.1063/1.5118360/suppl_file/movie_2.mp4}{Movie 2}).\\
%\begin{acknowledgement}%! achemso

\begin{acknowledgements} %! revtex4; sがつく。acknowledgements
%This work was supported by a grant for `JST-SENTAN' (to S.W.) and Grant for Young Scientists from Hokuriku Bank (to S.W.) and JSPS KAKENHI Grant Numbers JP26790048 (to S.W.), JP16H00799 (to S.W.), 17H04818 (to S.W.)and JP26119003 (to T.A.). This work was also supported by Kanazawa University CHOZEN PROJECT.
This work was supported by a grant of JST SENTAN (JPMJSN16B4 to S.W.), Grant for Young Scientists from Hokuriku Bank (to S.W.), JSPS Grant-in-Aid for Young Scientists (B) (JP26790048 to S.W.), JSPS Grant-in-Aid for Young Scientists (A) (JP17H04818 to S.W.), JSPS Grant-in-Aid for Scientific Research on Innovative Areas (JP16H00799 to S.W.) and JSPS Grant-in-Aid for Challenging Exploratory Research (JP18K19018 to S.W.) and JSPS Grant-in-Aid for Scientific Research (S) (JP17H06121 and JP24227005 to T.A.). This work was also supported by a Kanazawa University CHOZEN project and World Premier International Research Center Initiative (WPI), MEXT
, Japan.
%\end{acknowledgements}
\end{acknowledgements}

%%%%%%%%%%%%%%%%%%%%%%%%%%%%%%%%%%%%%%%%%%%%%%%%%%%%%%%%%%%%%%%%%%%%%
%% The same is true for Supplementary Information, which should use the
%\begin{suppinfo}

%This will usually read something like: ``Experimental procedures and
%characterization data for all new compounds. The class will
%automatically add a sentence pointing to the information on-line:
%\end{suppinfo}

%\bibliography{achemso-demo}
%\bibliography{D:/Dropbox/submission/Biblio/reference_20141021.bib}
%\bibliography{D:/Dropbox/submission/Biblio/reference_20141021.bib}
\bibliography{../Biblio/reference_20141021.bib}
%\bibliography{reference_20141021.bib}
\end{document}